\documentclass[preprint,showpacs,english,aps]{revtex4-1}
\usepackage{babel,rotating,dcolumn}
\usepackage{colordvi,graphicx,color,amsbsy,amsmath,bm,amssymb}
\usepackage{ulem} 
\usepackage{comment}
\begin{document}
\title[]
{Scale-dependent statistics of inertial particle distribution in high Reynolds number turbulence
}
\author{Keigo Matsuda$^{1}$} 
\email{k.matsuda@jamstec.go.jp}
\author{Kai Schneider$^{2}$} 
\author{Katsunori Yoshimatsu$^{3}$} 
\affiliation{$^{1}$ Research Institute for Value-Added-Information Generation (VAiG), Japan Agency for Marine-Earth Science and Technology (JAMSTEC), Yokohama 236-0001, Japan}
\affiliation{$^{2}$ Institut de Math\'ematiques de Marseille (I2M), Aix-Marseille Universit\'e, CNRS and Centrale Marseille, 39 rue F. Joliot-Curie, 13453 Marseille Cedex 13, France}
\affiliation{$^{3}$ Institute of Materials and Systems for Sustainability, Nagoya University, Nagoya, 464-8601, Japan}
\date{\today}
\begin{abstract}
Multiscale statistical analyses of inertial particle distributions are presented to investigate the statistical signature of clustering and void regions in particle-laden incompressible isotropic turbulence.
Three-dimensional direct numerical simulations of homogeneous isotropic turbulence at high Reynolds number ($Re_\lambda \gtrsim 200$) with up to $10^9$ inertial particles are performed
for Stokes numbers ranging from $0.05$ to $5.0$.
Orthogonal wavelet analysis is then applied to the computed particle number density fields. 
Scale-dependent skewness and flatness values of the particle number density distributions are calculated and the influence of Reynolds number $Re_\lambda$ and Stokes number $St$ is assessed. 
For $St \sim 1.0$, both the scale-dependent skewness and flatness values become larger as the scale decreases, suggesting intermittent clustering at small scales. 
For $St \le 0.2$, the flatness at intermediate scales, i.e. for scales larger than the Kolmogorov scale and smaller than the integral scale of the flow, increases as $St$ increases, and the skewness exhibits negative values at the intermediate scales.
The negative values of the skewness are attributed to void regions. 
These results indicate that void regions at the intermediate sales are pronounced and intermittently distributed for such small Stokes numbers.
As $Re_\lambda$ increases, the flatness increases slightly. 
For $Re_\lambda \ge 328$, the skewness shows negative values at large scales, 
suggesting that void regions are pronounced at large scales, while clusters are pronounced at small scales.
\end{abstract}

\pacs{47.27.Ak, 47.27.ek, 47.27.Gs}
\maketitle
\section{Introduction}

Inertial particles suspended in three-dimensional (3D) turbulent flows are ubiquitous in geophysical flows.
The spectrum of applications covers plankton dynamics, pollution dispersion in cities or in the atmosphere, or even the planet formation in the early age of our universe.
The precipitation mechanism in convective clouds, where inertial particles (i.e., water droplets) are suspended in high Reynolds number turbulence, is of particular interest in atmospheric flow~\cite{Shaw}. %
For instance cloud droplet motion in turbulence increases the collision coalescence frequency and enhances the rain drop formation. The importance of turbulence in the collision coalescence process is well summarized in the introduction of Ref.~\cite{Ireland2016}. 
One of the key factors that determines  the droplet collision coalescence frequency is turbulent clustering of cloud droplets.
When the particle size is smaller than the smallest turbulent scale, i.e., the Kolmogorov scale, and the particle density is larger than the fluid density, inertial particle motion deviates from turbulent flow motion, and particles form a nonuniform number density distribution, which consists of cluster (large number density) and void (small number density) regions. 
%
Clustering of cloud droplets can also increase the radar reflectivity factor~\cite{Matsuda2014,MO2019} due to the interference of microwaves scattered by spatially correlated droplets. Quantitative estimates of the increase in the radar reflectivity factor require the Fourier spectrum of number density fluctuations of turbulent clustering particles which covers scales comparable to radar wavelengths. 
Sound modeling becomes necessary for improving weather prediction and requires thus deep insight into the nonlinear multiscale dynamics.  



Inertial particle clustering in homogeneous isotropic turbulence was investigated in many publications.
For review articles on this topic, we refer readers to, e.g., Refs.~\cite{Toschi,Monchaux2012}.
In this paper, we consider inertial heavy particles; i.e., the particle density is sufficiently larger than the fluid density. The particle acceleration balances the drag force, which is proportional to the velocity difference of a particle and fluid, and the proportional coefficient is given by the inverse of the particle relaxation time, $\tau_p$.
The dimensionless parameter for $\tau_p$ is the Stokes number, which is defined as $St \equiv \tau_p/\tau_\eta$, where $\tau_\eta$ is the Kolmogorov time.
The clustering of inertial heavy particles was first explained by the preferential concentration mechanism~\cite{Maxey,Squires}, in which inertial particles are swept out from strong vortices due to centrifugal effects and concentrate in low-vorticity and high-strain-rate regions when the particle relaxation time is sufficiently small compared to the time scale of vortices.
%
%
The pair correlation function (PCF) is widely used to analyze clustering, because it is directly related with the particle collision rate~\cite{Sundaram_Collins1997,Wang2000}.
%
The PCF shows typically a power-law behavior at sub-Kolmogorov scales and the slope 
is dependent on the Stokes number $St$.
%
Boffetta {\it et al}.~\cite{Boffetta2004} pointed out that a multiscale structure of clustering can be observed  
in the inverse cascade range in two-dimensional turbulence.
They showed that the probability density function (PDF) of void area exhibits a power-law, independent of the Stokes number.
Yoshimoto \& Goto \cite{YG2007} reported similar results for the PDF of void volumes at scales larger than the Kolmogorov scale in homogeneous isotropic turbulence using 3D direct numerical simulation (DNS). 
%
Coleman \& Vassilicos \cite{CV2009} further 
showed that the scale similarity of particle distribution is explained by the sweep-stick mechanism proposed by Goto \& Vassilicos \cite{GV2006}, in which particles are swept by large-scale flow motion while sticking to stagnation points of Lagrangian fluid acceleration (see also Refs. \cite{CGV2006,GV2008}).
The multiscale structure of clustering was also observed in experiments by Monchaux {\it et al}.~\cite{Monchaux2010}: They measured particle distribution in a wind tunnel and reported that both PDFs of void and cluster areas exhibit power-laws independent of the Stokes number. 
%
Bec {\it et al.}~\cite{Bec2007} discussed the scale dependence of particle distribution, using the PDF of particle mass density, coarse grained on scales in the inertial range based on their 3D DNS data. They reported that the PDF is changing with the scale-dependent contraction rate. 
%
The multiscale clustering structure was also analyzed using the PCF~\cite{YG2007,Saw2008} 
and the Fourier spectrum of number density fluctuations~\cite{Matsuda2014}.
It was shown that both PCF and Fourier spectrum are strongly dependent on the Stokes numbers even at scales larger than the Kolmogorov scale.
The scale similarity of particle clustering in the inertial range of turbulence was also discussed on the basis of theoretical analyses (e.g., Refs. \cite{Bragg2015, Ariki2018}).
However, the multiscale clustering structure is not fully described by such theoretical analyses.
Bassenne {\it et al}. \cite{Bassenne2017} proposed a wavelet-based method to extract coherent clusters of inertial particles in fully developed turbulence. 
Wavelet multiresolution statistics of particle-laden turbulence has been recently introduced in Ref. \cite{Bassenne2018} for studying
the cross-correlations between energy spectra of the fluid and the dispersed-phase field variables in particle-laden turbulence.
Wavelets represent turbulent flow fields in scale and position, complementary to Fourier techniques which yield insight into wave number contributions of turbulent flows.
Hence the wavelet representation can quantify spatial fluctuations at different scales, which is a key for analyzing spatial intermittency. This is possible due to the local and oscillatory character of the wavelet basis functions which yield an efficient orthogonal representation of the flow field thanks to fast algorithms.
For the Fourier transform this task is out of reach owing to the global character of the basis functions.
Wavelet techniques for turbulent flow have already some history starting with the work of e.g., Refs. \cite{FS89, Fa92,Mene91}. 
Numerous applications can be found to extract coherent vorticity \cite{FSK99,FPS01,OYSFK07}, quantifying intermittency \cite{SFK04, BLS07}, performing scale-dependent statistics \cite{YOSKF09} and turbulence modeling \cite{FS01,SFPR05}. A review for computing turbulent flows can be found in Ref.~\cite{SV10}.
Recently, orthogonal wavelets have been applied to active matter turbulence \cite{activematter}, turbulent premixed combustion \cite{combust} and droplet-laden turbulence \cite{dropwave}. 

%
The aim of the current work is to get insight into the scale-dependent statistics of the particle distribution and into the multiscale structure of clusters and voids in particle-laden turbulence.
To this end orthogonal wavelet decomposition of the particle number density fields is performed.
The analyzed data are obtained by DNS of 3D homogeneous isotropic turbulence at high Reynolds number laden with inertial particles, where the Taylor-microscale based Reynolds number is $Re_\lambda \gtrsim 200$. 
The influence of different physical parameters, Reynolds number $Re_\lambda$ and Stokes number $St$,
is assessed.

The remainder of the paper is organized as follows.
First, we briefly  summarize the governing equations and the performed DNS computations in Sec.~\ref{sec2}.
In Sec.~\ref{sec3}, we describe the wavelet methodology and wavelet-based statistical measures to quantify the scale-dependent distribution of the particle number density field.
Numerical results are then presented in~Sec. \ref{sec4}. 
Finally, Sec.~\ref{sec5} draws some conclusions and gives perspectives for future work. 

\section{Particle-laden turbulence}\label{sec2}
We present the governing equations of particle-laden turbulence in Sec.~\ref{secBE}, and describe the DNS computations  in Sec.~\ref{secDNS}.
In Sec.~\ref{secNDF} we explain the conversion of the Lagrangian particle data into an Eulerian number density field, including its Fourier spectrum.
\subsection{Basic equations} \label{secBE}
We consider a homogeneous velocity field ${\bm u}({\bm x},t)$ of an incompressible fluid obeying the Navier--Stokes equation together with the divergence free condition:
\begin{eqnarray}
\frac{\partial {\bm u} }{\partial t} + ({\bm u} \cdot \nabla) {\bm u}  &=& -\frac{1}{\rho} \nabla p +\nu \nabla^2 {\bm u} + {\bm f},\label{NS_eq2} \\
\nabla \cdot {\bm u} &=& 0, \label{NS_eq}
\end{eqnarray}
where ${\bm x}=(x_1,x_2,x_3)$, $\nabla=(\partial/\partial x_1, \partial/\partial x_2, \partial/\partial x_3)$, 
$t$ is the time, 
${\bm f}({\bm x},t)$ is an external solenoidal forcing, 
$p({\bm x},t)$ is the pressure,
$\nu$ is the kinematic viscosity of the fluid, and $\rho$ is the density.
The equations are completed with periodic boundary conditions and a suitable initial condition.
Here and in the following, we omit the arguments ${\bm x}$ and $t$, unless otherwise stated.


We assume that the particle size is sufficiently smaller than the Kolmogorov scale 
and the particle density $\rho_p$ is sufficiently larger than the fluid density $\rho$ (i.e., $\rho_p/\rho \gg 1$). 
Then, Lagrangian motion of inertial heavy particles can be described by 
\begin{eqnarray}
    \frac{d {\bm x}_p}{d t} &=& {\bm v}
    \label{eq:particle_pos}, \\
    \frac{d {\bm v}}{d t} &=& -\frac{{\bm v} - {\bm u}}{\tau_p} 
    \label{eq:particle_vel},
\end{eqnarray}
where ${\bm x}_p$ and ${\bm v}$ are the position and velocity of a Lagrangian particle, and $\tau_p$ is the relaxation time of particle motion.
Assuming the Stokes flow for spherical particles, $\tau_p$ is given by
\begin{equation}
    \tau_p = \frac{\rho_p}{\rho} \frac{2 a^2}{9 \nu}
    \label{eq:taup},
\end{equation}
where $a$ is the particle radius.

The important parameters in this study are the Taylor-microscale based Reynolds number $Re_\lambda$ and the Stokes number $St$.
The Taylor-microscale based Reynolds number $Re_\lambda$ is defined as $Re_\lambda \equiv u'\lambda/\nu$, where $u'$ is the turbulent velocity fluctuation $u' \equiv \sqrt{\langle|{\bm u}|^2\rangle/3}$, and $\lambda$ is the Taylor microscale $\lambda \equiv \sqrt{15 \nu u'^2/\epsilon}$. $\epsilon$ is the energy dissipation rate, defined by $\epsilon \equiv \nu \left\langle\frac{\partial u_i}{\partial x_j}\frac{\partial u_i}{\partial x_j}\right\rangle$, 
and $\langle\cdot\rangle$ denotes an ensemble average.
The Stokes number $St$ indicates the contribution of particle inertia and defined as $St \equiv \tau_p/\tau_\eta$, where $\tau_\eta$ is the Kolmogorov time ($\tau_\eta \equiv \sqrt{\nu/\epsilon}$).
In homogeneous turbulence the ensemble average can be regarded as space and time average under appropriate assumptions.



\subsection{Direct numerical simulation}\label{secDNS}
The DNS of particle-laden turbulence was performed using the same DNS program as that used in Ref. \cite{Matsuda2014}.
Equations~(\ref{NS_eq2}) and (\ref{NS_eq}) were solved on Cartesian staggered grids. 
The fourth-order central-difference schemes were used for the advection and viscous terms \cite{Morinishi1998} and the second-order Runge--Kutta scheme was used for time integration. The velocity and pressure were coupled by the highly simplified marker and cell (HSMAC) method \cite{Hirt&Cook1972}, where the second-order central difference scheme was used for the pressure gradient. 
To obtain statistically steady-state turbulence, a parallelized external solenoidal forcing \cite{Onishi2011} was applied to the large scales satisfying $k < 2.5$. Here $k=|{\bm k}|$ is a magnitude of wave number vector ${\bm k}$.
Equations~(\ref{eq:particle_pos}) and (\ref{eq:particle_vel}) were solved for discrete Lagrangian points. The time integration scheme was the same as that for the flow field.


The computational cubic domain has side length of $2\pi$.
Periodic boundary conditions  are applied in $x_1$, $x_2$ and $x_3$ directions. 
The domain was discretized uniformly into $N_{grid}^3$ grid points, giving a grid spacing of $\Delta = 2\pi/N_{grid}$.
The DNS was performed for three turbulent flows at different Reynolds numbers; Flow 1, Flow 2 and Flow 3. 
The resolution was chosen to satisfy $k_{max}\eta \approx 2$, where $k_{max}$ is the maximum wave number given by $k_{max}=\pi/\Delta$, and $\eta = (\nu^3/\epsilon)^{1/4}$ is the Kolmogorov scale.
Inertial particles were imposed uniformly and randomly in the computational domain at $t=0$, where the turbulent flow field had reached a statistically steady state. 
Particle position data were sampled at 10 time instance of $t=11 T_0$ to $20 T_0$ at interval of $T_0$, where $T_0$ is the dimensionless time unit and comparable to the eddy-turnover time. 
The Stokes number $St$ of inertial particles was set to 0.05, 0.1, 0.2, 0.5, 1.0, 2.0 and 5.0 for Flow 1,
and the particle motion of $St=1.0$ was tracked for Flow 2 and Flow 3.
The statistics of the obtained turbulent flows and the number of particles $N_p$ are summarized in Table~\ref{tab:DNS}.
Note that $N_p$ particles were tracked for each case of $St$ for Flow 1.
Time average for the statistics was taken for the period of $10T_0 \le t \le 20T_0$.







\begin{table}[t!]
\begin{center}
\begin{tabular}{c|cccccc}\hline\hline
      &\,\, $N_{grid}$ \,\, &\,\, $Re$ \,\,& \,\,$Re_\lambda$ \,\,& \,\,$u'$\,\, & \,\,$k_{max}\eta$\,\, & \,\,$N_p$\,\,              \\\hline
Flow 1 \,\,& 512        & 909  & 204         & 1.01 & 2.02          & 1.07 $\times 10^9$ \\
Flow 2  \,\,& 1024       & 2220 & 328         & 1.00 & 2.12          & 5.00 $\times 10^7$ \\
Flow 3  \,\,& 2048       & 5595 & 531         & 1.00 & 2.14          & 4.00 $\times 10^8$ \\\hline\hline
\end{tabular}
\end{center}
\caption{DNS parameters and statistics of obtained turbulence; the number of grid points $N_{grid}$, the Reynolds number of DNS $Re=\nu^{-1}$, the Taylor-microscale based Reynolds number $Re_\lambda$, the turbulent velocity fluctuation $u'$, $k_{max} \eta$, and the number of particles $N_p$. 
}
\label{tab:DNS}
\end{table}

\subsection{Number density fluctuations}\label{secNDF}




The number density field of the discrete particle positions can be described as
\begin{equation}
    n_{\delta}({\bm x},t) = \frac{1}{n_0} \, \sum_{m=1}^{N_p} \delta \left({\bm x} - {\bm x}_{p,m}(t) \right)
    \label{eq:n_delta},
\end{equation}
where $\delta({\bm x})$ is the Dirac delta function, the subscript $m$ denotes the identification number of the particle, and $n_0$ is the scaling factor: The mean dimensional number density $n_0 = N_p/(2\pi)^3$ is used 
in order that $\langle n_\delta \rangle = 1$.
However, wavelet analysis cannot be applied directly to $n_{\delta}({\bm x},t)$. 
Thus, to apply the wavelet analysis, the number density field in Eq.~(\ref{eq:n_delta}) was converted to the number density field data on equidistant grid points based on the histogram method; i.e., the computational domain was discretized into an array of $N_g^3$ equally sized boxes, and the number of particles in each box was counted. 
%
%
The histogram method, which corresponds to the zeroth-order kernel density estimation, retains fine clustering structures better than higher-order kernels.
The number density field based on the histogram method is given by
\begin{equation}
    n({\bm x},t) \, = \, \sum_{i_1,i_2,i_3=0}^{N_g-1}
        \left\{ \int_\mathbb{T} K_h({\bm x}_{i_1,i_2,i_3} - {\bm x}')n_\delta({\bm x}',t) d{\bm x}' \right\} 
        h^3 K_h({\bm x} - {\bm x}_{i_1,i_2,i_3})
    \label{eq:particlenumberfield},
\end{equation}
where $\mathbb{T} = 2 \pi \mathbb{R} / {\mathbb{Z}}$, 
${\bm x}_{i_1,i_2,i_3}$ is the box position given by ${\bm x}_{i_1,i_2,i_3} = h(i_1+1/2, i_2+1/2, i_3+1/2)$, and $K_h({\bm x})$ is a piecewise constant function defined as $K_h({\bm x})=1/h^3$ for $-h/2 \le x_i < h/2$ ($i=1, 2, 3$), 
while $K_h({\bm x})=0$ otherwise. 
Here $h$ denotes the width of the piecewise function, and for the histogram we have $h=2\pi/N_g$.
Note that Eq.~(\ref{eq:particlenumberfield}) satisfies $\langle n \rangle = 1$.
%
For the number density field $n({\bm x},t)$ the number of grid points in each direction was set to $N_g = 1024$, independently of the number of grid points $N_{grid}$ in the DNS.
The influence of $N_g$ on the wavelet-based statistics is discussed in Appendix \ref{gnpn}.
Bassenne {\it et al}. \cite{Bassenne2017} also used the histogram method to obtain the number density field for the wavelet analysis. 
Nguyen {\it et al}. \cite{Romain2010} used the kernel density estimation with the Gaussian kernel, but the Gaussian kernel smooths out fine clustering structures because it works as a blunt low-pass filter.
\section{Wavelet analysis of the number density field}\label{sec3}

The scale-dependent statistics of the particle number density field $n({\bm x},t)$, Eq. (\ref{eq:particlenumberfield}), are based on an orthogonal wavelet decomposition which is summarized in Sec. \ref{secOWD}. For details on wavelets we refer the reader to textbooks, e.g. Refs. \cite{Daubechies1992, Mallat98}. The scale-dependent moments of the number density field yield statistical estimators of the different quantities considered such as variance, skewness and flatness values, and are defined in Sec. \ref{secWS}.
\subsection{Orthogonal wavelet decomposition}\label{secOWD}
We consider here a scalar field $n({\bm x}, t)$, i.e., the particle number density field at a given instant $t$, in the $(2\pi)^3$ periodic cube.
The field is decomposed into a 3D orthogonal wavelet series, and it is thus unfolded into scale, positions and seven directions ($\mu = 1, \cdots, 7$).
The 3D mother wavelet $\psi_\mu({\bm x})$ is hereby based on a tensor product construction and a family of wavelets 
%
%
$\psi_{\mu,{\bm \gamma}} ({\bm  x})$ can be generated by dilation and translation. This family yields an orthogonal basis of $L^2({\mathbb{R}}^3)$.
The multi-index ${\bm \gamma}= (j,i_1, i_2, i_3)$ denotes  the scale $2^{-j}$ and  position $2\pi \times 2^{-j} {\bm i} =2\pi \times 2^{-j}(i_1, i_2, i_3)$ of the wavelets for each direction, and $i_\ell=0,\cdots,2^j-1$ $(\ell=1,2,3)$.
The wavelets are well-localized in space around position $2 \pi \times  2^{-j}{\bm i}$ and scale $2^{-j}$, oscillating, and smooth.
Application of a periodization technique \cite{Mallat98} to the wavelets generates likewise an orthogonal basis of $L^2({\mathbb{T}}^3)$.
%
The spatial average of $\psi_{\mu,{\bm \gamma}} (\bm x)$, defined by $\langle \psi_{\mu,{\bm \gamma}} \rangle =(2\pi)^{-3} \int_{\mathbb{T}} \psi_{\mu,{\bm \gamma}}({\bm x}) d{\bm x}$, vanishes for each index, which is a necessary condition for being a wavelet.

The number density field $n(\bm x)$ sampled on $N_g^3=2^{3J}$ equidistant grid points, can be developed into an orthogonal wavelet series: 
\begin{equation}
n({\bm  x}) \, = \, \langle n({\bm x}) \rangle \, + \, 
\sum_{j=0}^{J-1} n_j({\bm  x}), 
\label{OWS}
\end{equation}
where $n_j({\bm x})$ is the contribution of $n({\bm x})$ at scale $2^{-j}$ defined by 
\begin{equation}
n_j({\bm  x}) = \sum_{\mu=1}^7 \sum_{i_1,i_2,i_3 =0}^{2^j -1} {\widetilde n}_{\mu, {\bm \gamma}}  \psi_{\mu, {\bm \gamma}} ({\bm  x}),
\label{OWS_2}
\end{equation}
and $\langle n({\bm x}) \rangle$ is the mean value.
Due to orthogonality of the wavelets, the coefficients are given by $ {\widetilde n}_{\mu,{\bm \gamma}}  = \left< n, \psi_{\mu,{\bm \gamma}} \right>$, where $\left<\cdot,\cdot\right>$ denotes the $L^2$-inner product defined by $\left< \xi, \zeta \right> = (2\pi)^{-3} \int_{\mathbb{T}} \xi({\bm  x}) \, \zeta({\bm  x}) d{\bm  x}$. 
At scale $2^{-j}$ we have $7 \times 2^{3j}$ wavelet coefficients for $n({\bm  x})$.
Thus, in total we have  $N_g^3$ coefficients for each component of the vector field corresponding to $N_g^3-1$ 
wavelet coefficients and the non-vanishing mean value. 
These coefficients are efficiently computed from the $N_g^3$ grid point values for $n({\bm x})$ using the fast wavelet transform, which has linear computational complexity.


The scale $2^{-j}$ of the wavelet transform  and the wave number $k_j$ of the Fourier transform
are related via
\begin{equation}
k_j = k_\psi 2^j, \label{coifkj}
\end{equation} 
where $k_\psi$ is the centroid wave number of the chosen wavelet. 
For the Coiflet 12 wavelet chosen here, which has four vanishing moments, we have $k_\psi = 0.77$.

\subsection{Wavelet-based statistics of the particle number density field}\label{secWS}
We discuss scale-dependent statistics of the particle number density field $n({\bm x})$ which are based on scale-dependent moments using the wavelet decomposition of Eq.~(\ref{OWS}) 
We define the $q$-th order moments of $n_j({\bm x})$,
\begin{equation}
M_q [ n_j ] = \langle ( n_j )^q \rangle, 
\end{equation}
and note that by construction the mean value vanishes, $\langle n_j \rangle=0$. 
The moments are thus central moments.
These scale-dependent moments are intimately related to the $q$-th order structure functions~\cite{SFK04}.

In the following, we consider the second order moment $M_2[n_j ]$,
the third order moment $M_3[n_j ]$, and the fourth order moment $M_4[n_j ]$.
%
The wavelet energy spectrum of $n_j({\bm x})$ can be defined using the second order moment $M_2[n_j]$ and Eq. (\ref{coifkj}),
\begin{equation}
{E} [n_j]  = \frac{1}{\Delta k_j} M_2 [ n_j ] ,
\label{wave_spe}
\end{equation}
where $\Delta k_j = (k_{j+1} - k_{j}) \ln 2$ \cite{Mene91}.
The wavelet spectrum $E[n_j]$ corresponds to a smoothed version of the Fourier energy spectrum~\cite{Fa92, Mene91}.
The orthogonality of the wavelets implies that we obtain the variance of the number density field $\sum_{j=0}^{J-1} {E}[n_j] $.
The asymmetry of the PDF 
of $n_j({\bm x})$ can be quantified by its skewness defined as
\begin{equation}
S [n_j ]=\frac{M_3 [ n_j]}{ \left(M_2 [ n_j  ] \right)^{3/2} }.\label{skewness}
\end{equation}
The scale-dependent flatness, which measures the intermittency at scale $2^{-j}$, is defined by
\begin{equation}
F [n_j ]=\frac{M_4 [ n_j]}{ \left(M_2 [ n_j  ] \right)^2 }.\label{flatness}
\end{equation}
For a Gaussian distribution the flatness equals three at all scales.

In Ref. \cite{BLS07} it was shown that the flatness is directly related to the energy spectrum of Eq.~(\ref{wave_spe}) and the standard deviation of the spatial variability of $E[n_j]$,
\begin{equation}
F[n_j ] = \left( \frac{{\vartheta} [n_j]}{ {E} [n_j ] } \right)^2 \, + \, 1, \label{flatness_std}
\end{equation}
where ${\vartheta} [n_j]$ is the standard deviation and defined as $ {\vartheta} [n_j] = (1/\Delta k_j) \sqrt{ M_4 [ n_j ]  - \left( M_2 [ n_j ] \right)^2} $. This relation illustrates that the spatial variability of the spectrum, quantified by the fourth order moment, is reflected in increasing flatness values at small scales.

\section{Numerical results}\label{sec4}

\subsection{Scale-dependence of particle clusters and voids}
\begin{figure*}[tb!]
    \centering
    \begin{tabular}{ll}
        (a) & (b) \\
        \includegraphics[width=8cm]{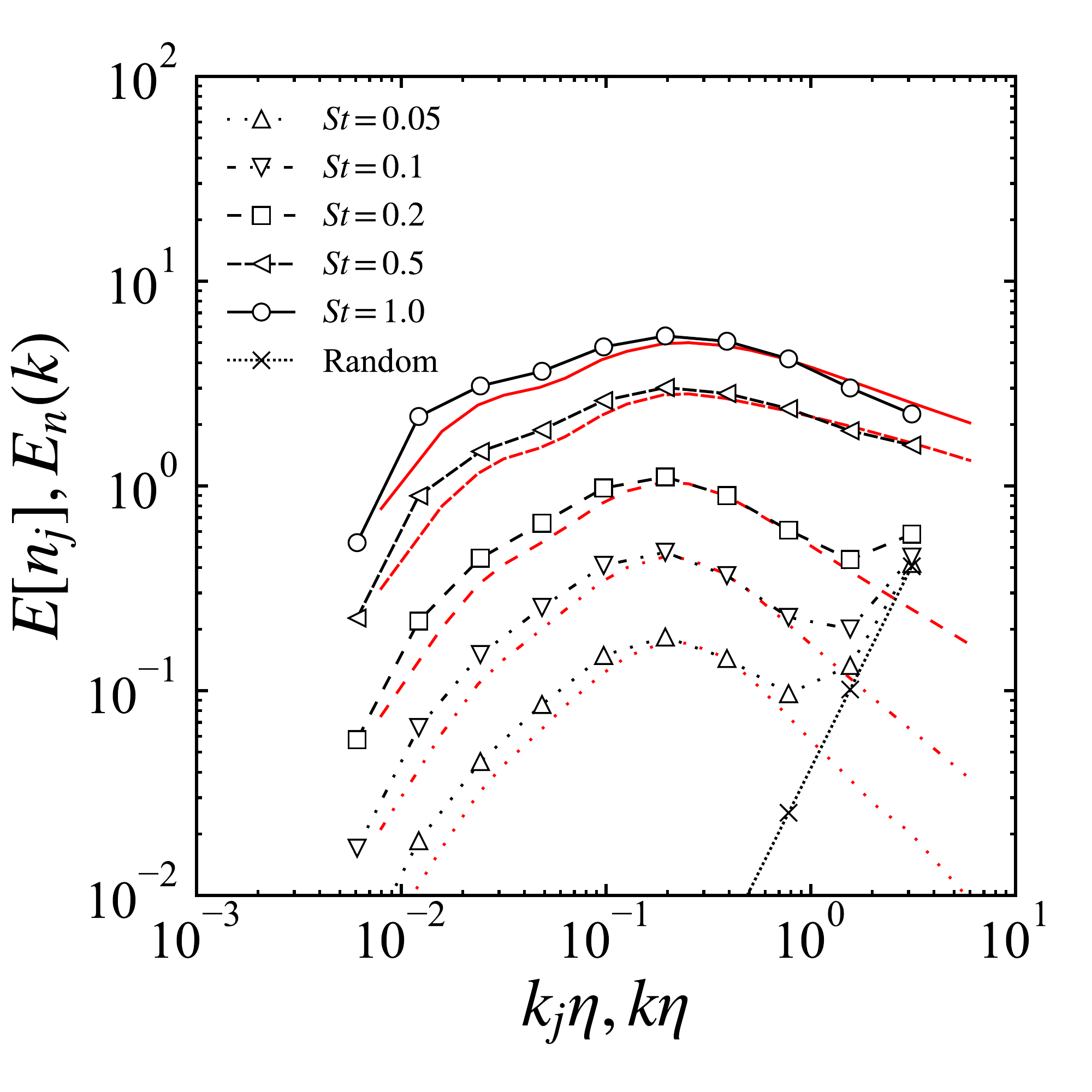} & 
        \includegraphics[width=8cm]{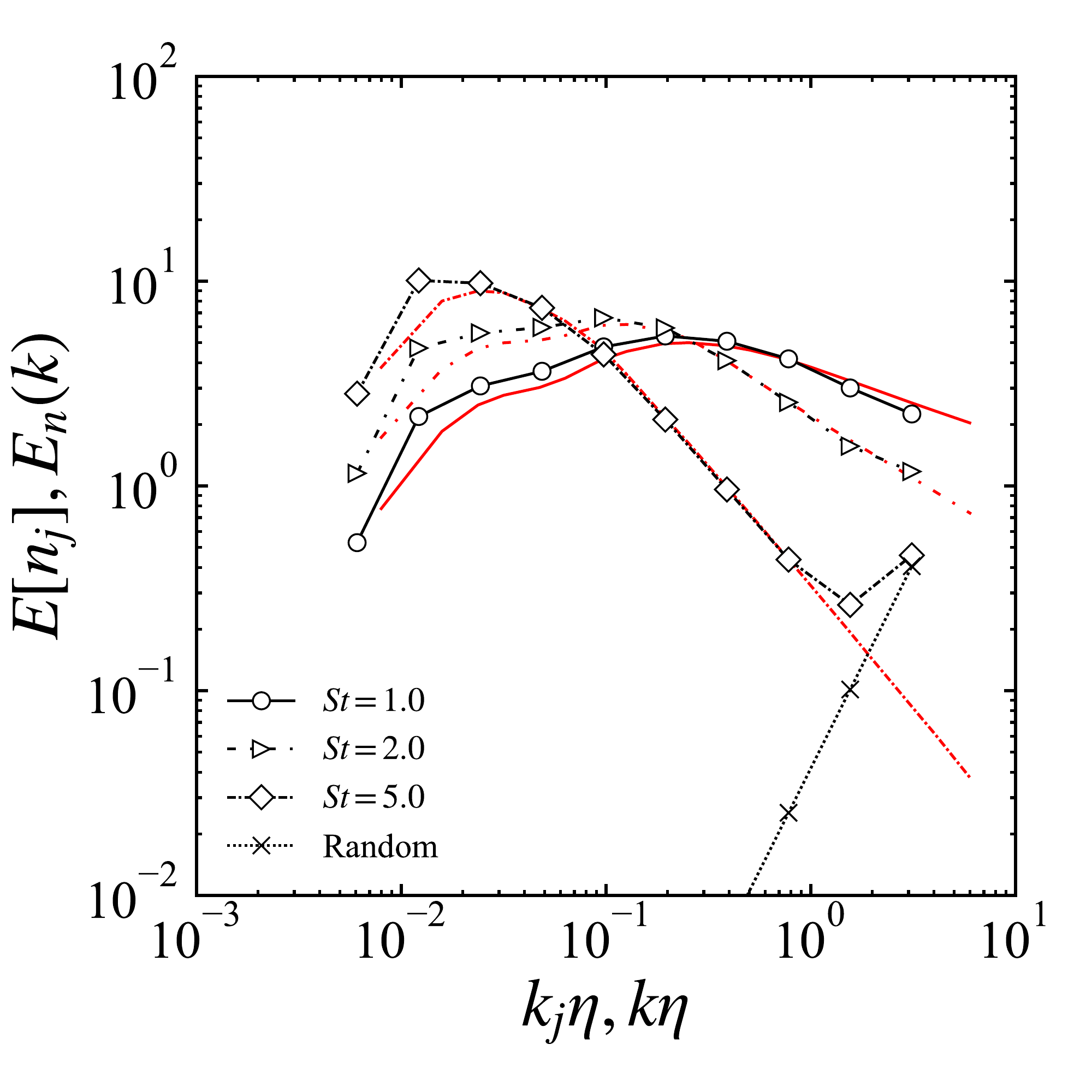}
    \end{tabular}
    \caption{Wavelet spectra $E[n_j]$ (black) and Fourier spectra $E_n(k)$ (red) of number density fluctuation at $Re_\lambda=204$ 
    for the cases of (a) $St \le 1$ and (b) $St \ge 1$. }
    \label{fig:Ej_St}
\end{figure*}
Figure~\ref{fig:Ej_St} presents wavelet spectra of number density fluctuations $E[n_j]$ together with number density Fourier spectra $E_n(k)$ at different Stokes numbers. 
The wave numbers $k_j$ and $k$ are normalized by the Kolmogorov scale $\eta$.
In Fig.~\ref{fig:Ej_St}(a), we can see that the spectra $E[n_j]$ increase with $St$ for each $k_j \eta$ when $St \le 1.0$.
This increase suggests that the particle clustering becomes prominent as $St$ becomes larger.
In the case that $St \ge 1.0$,
at larger scales $k_j \eta \gtrsim 10^{-1}$ the spectra become larger with $St$ (see Fig.~\ref{fig:Ej_St}(b)).
In contrast, 
at scales satisfying $k_j \eta \lesssim  10^{-1}$, the spectra become smaller for each $k_j \eta$, as $St$ increases from unity.
This non-monotonic behavior of $E[n_j]$ in terms of $St$ shows that the scale of the most intense  particle clustering becomes larger with $St (\ge 1.0)$.
The $St$ dependence of $E[n_j]$ is in accordance with that reported by Ref.~\cite{Matsuda2014}.
We can also see that for each $St$, $E[n_j]$ is in good agreement with the number density Fourier spectra $E_n(k)$.
It should be noted that a number density Fourier spectrum could contain the Poisson noise caused by the discrete nature of particle distribution when the standard Fourier transform is applied to the number density field \cite{SaitoGotoh2018}. In the Fourier spectra in Fig.~1, the Poisson noise is removed by
using the analytical Fourier transform technique of Ref.~\cite{Matsuda2014}.
In contrast, the influence of the noise remains in the wavelet spectra and is observed for $St=0.05$, 0.1, 0.2 and 5.0.
The wavelet spectra $E[n_j]$ are also plotted in the case of random particle positions with uniform probability as reference: $E[n_j] \propto k_j^{2}$, 
where the PDF of particle number density satisfies the Poisson distribution. 

\begin{figure*}[p!]
    \centering
\includegraphics[height=21.0cm]{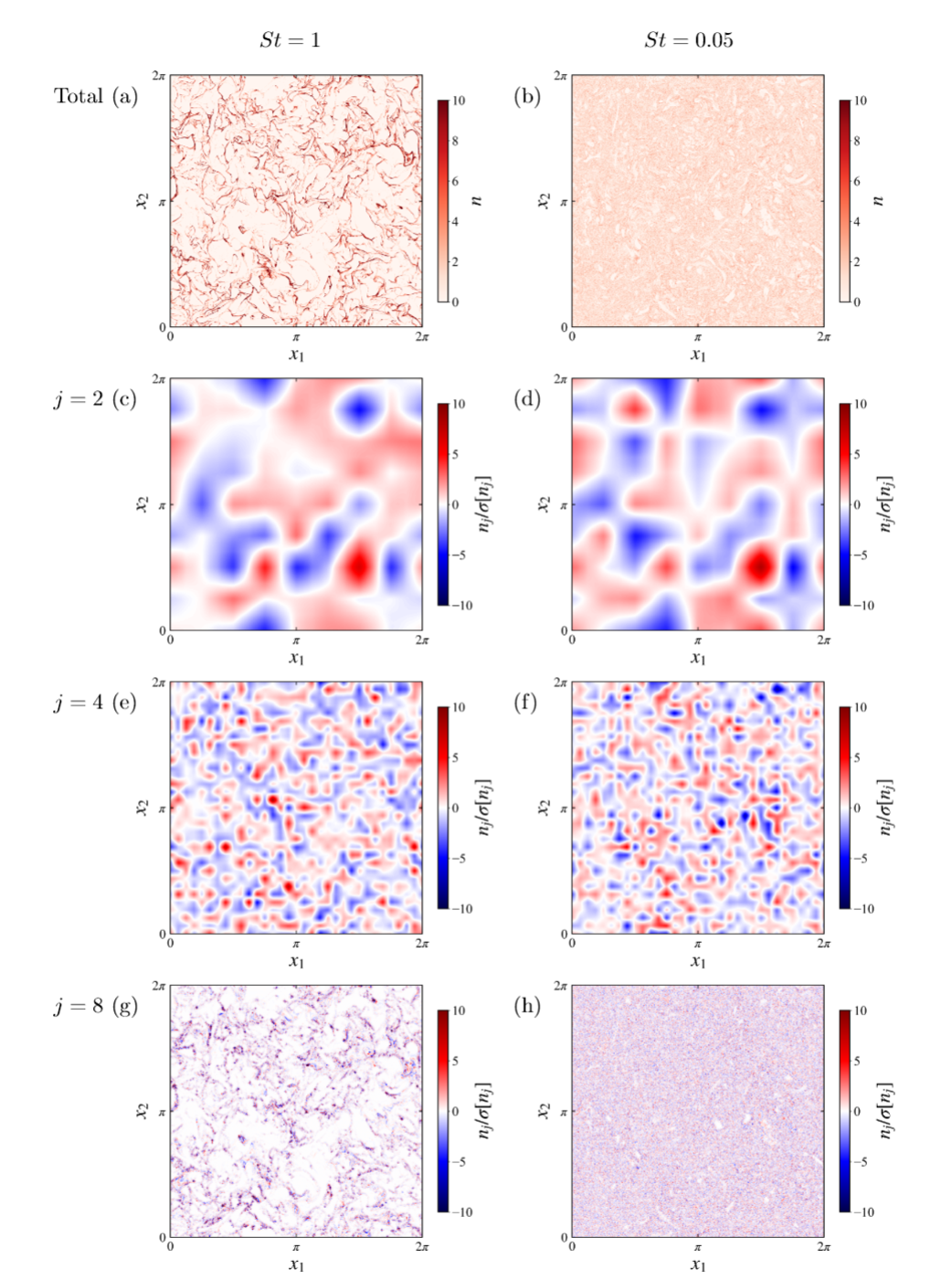}
\caption{Spatial distributions of total number density field $n({\bm x})$ (a, b) and scale contributions $n_j({\bm x})$ at $j=2$ (c, d), $j=4$ (e, f) and $j=8$ (g, h) in a $x_1$-$x_2$ cross section; (a, c, e, g) $St=1.0$, (b, d, f, h) $St=0.05$.
}
    \label{fig:density2d}
\end{figure*}

To get intuitive ideas about 
particle clustering and its scale dependence,
we visualize spatial distributions of scale-dependent number density fields $n_j$ on a two-dimensional plane at different scales for $St=1.0$ and $St=0.05$ together with the total number density fields $n({\bm x})$ in Fig.~\ref{fig:density2d}. 
The scale-dependent number density field $n_j$ is normalized by $\sigma[n_j]$, the standard deviation of $n_j$. Here, $\sigma[n_j]=\sqrt{M_2[n_j]}$.
In Figs.~\ref{fig:density2d}(a) and \ref{fig:density2d}(b), 
we can see 
the prominence of the particle clusters and void regions, especially at $St=1.0$.
The prominence of the clusters becomes substantial as scales becomes smaller, i.e., the scale index $j$ becomes larger.
In addition, it seems that the clusters and the voids are distributed more intermittently in space with increasing $j$ for each $St$.

\subsection{Reynolds number dependence}

\begin{figure*}[tb!]
    \centering
    \begin{tabular}{ll}
        (a) & (b) \\
        \includegraphics[width=8cm]{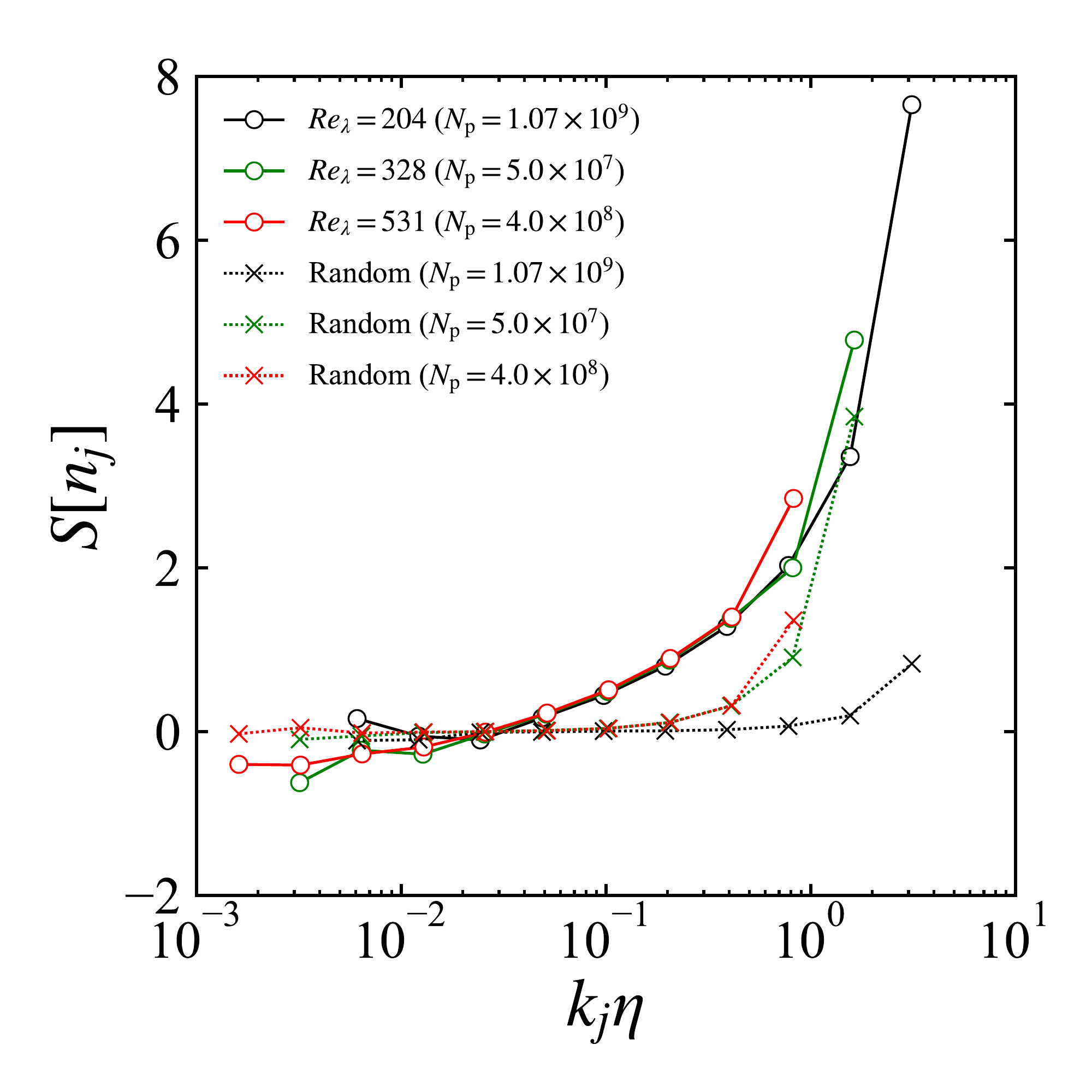} & 
        \includegraphics[width=8cm]{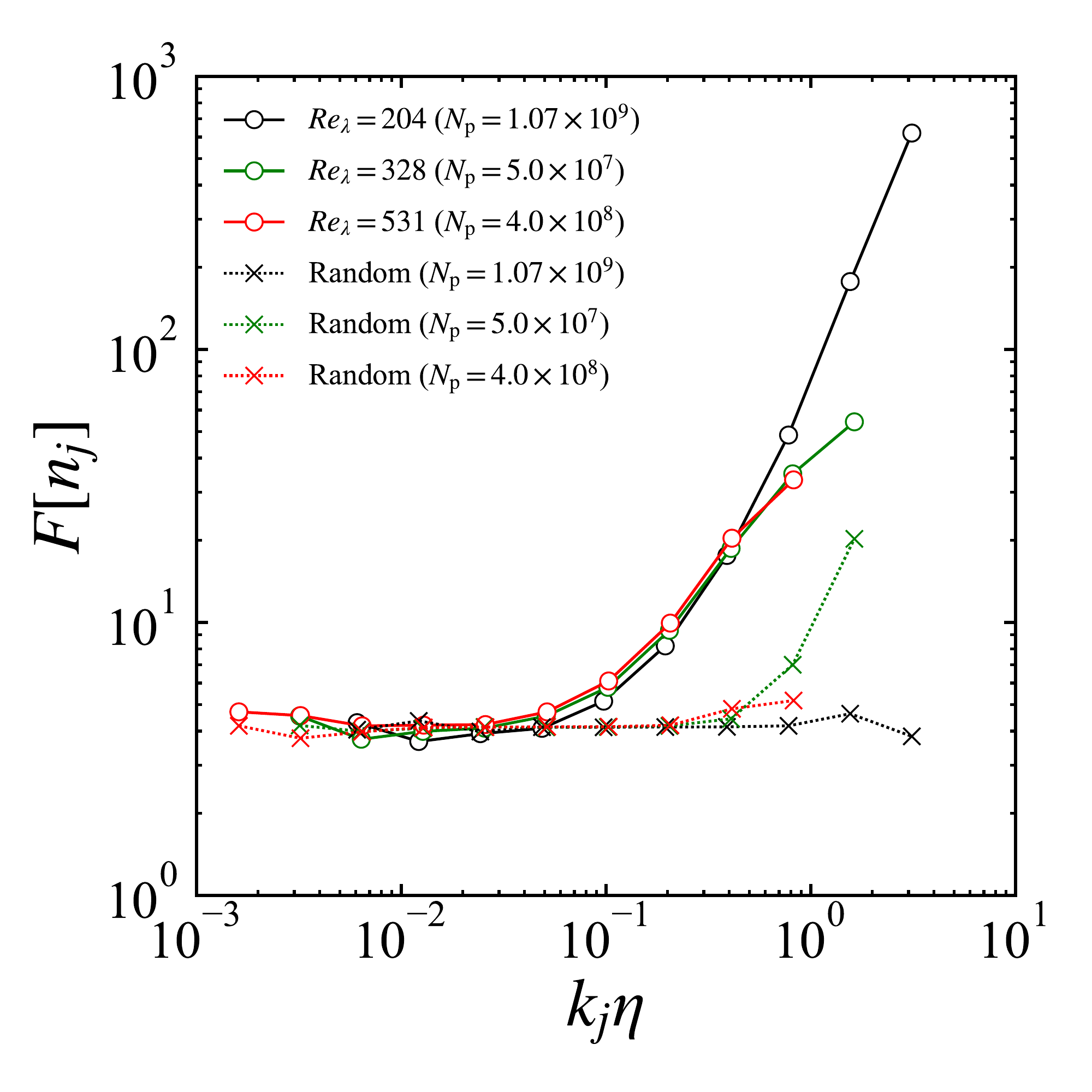}
    \end{tabular}
    \caption{Reynolds number dependence of scale-dependent skewness $S[n_j]$ and flatness $F[n_j]$ for $St=1.0$.}
    \label{fig:SjFj_Re}
\end{figure*}

We examine the influence of the Reynolds number $Re_\lambda$
on the scale-dependent skewness and flatness values of the particle number density, $S[n_j]$ and $F[n_j]$, for inertial particles at $St=1.0$ and those for randomly distributed particles.
The DNS for the three Reynolds numbers, $Re_\lambda$, uses
different number of particles (imposed by their computational cost), as shown in Table~\ref{tab:DNS}.
Thus, we also consider three sets of randomly distributed particles with the corresponding number of particles. 
%
Figure~\ref{fig:SjFj_Re} shows that skewness $S[n_j]$ and flatness $F[n_j]$ increase with $k_j \eta$, irrespective of the values of $Re_\lambda$. 
In Fig.~\ref{fig:SjFj_Re}(a), we can see that the skewness values $S[n_j]$ for three $Re_\lambda$ well collapse in the range $0.02 \lesssim k_j \eta \lesssim 0.5$, which suggests the $Re_\lambda$ dependence of $S[n_j]$ is negligible in  this $k_j \eta$ range. 
In contrast, Fig.~\ref{fig:SjFj_Re}(b) shows that $F[n_j]$ increases with $Re_\lambda$ for fixed $k_j \eta$ in the same range. 


Note that the statistics of the particle number density field for randomly distributed particles are equivalent to those for fluid particles $(St=0)$. The number density of the fluid particles is uniform due to the volume preserving nature of the incompressible flow. 
Thus void and clusters regions are absent and consequently the skewness values vanish and flatness values remain constant for the case of random particles, if the number of particles $N_p$ is sufficiently large.
This is confirmed in Appendix~\ref{gnpn}. 
Thus,  for $Re_\lambda=204$, $S[n_j]$ and $F[n_j]$ are nearly independent of $N_p$. 
%
For the higher Reynolds number cases, as illustrated by the randomly distributed particles in Fig.~\ref{fig:SjFj_Re}, the $N_p$ dependence of $S[n_j]$ and $F[n_j]$ is not quantitatively negligible for smaller scales $k_j \eta \gtrsim 0.5$. 
Thus, here we limit the discussion of the $Re_\lambda$ dependence only for $k_j \eta \lesssim 0.5$.
%

\subsection{Stokes number dependence}

\begin{figure*}[tb!]
    \centering
    \begin{tabular}{ll}
        (a) & (b) \\
        \includegraphics[width=8cm]{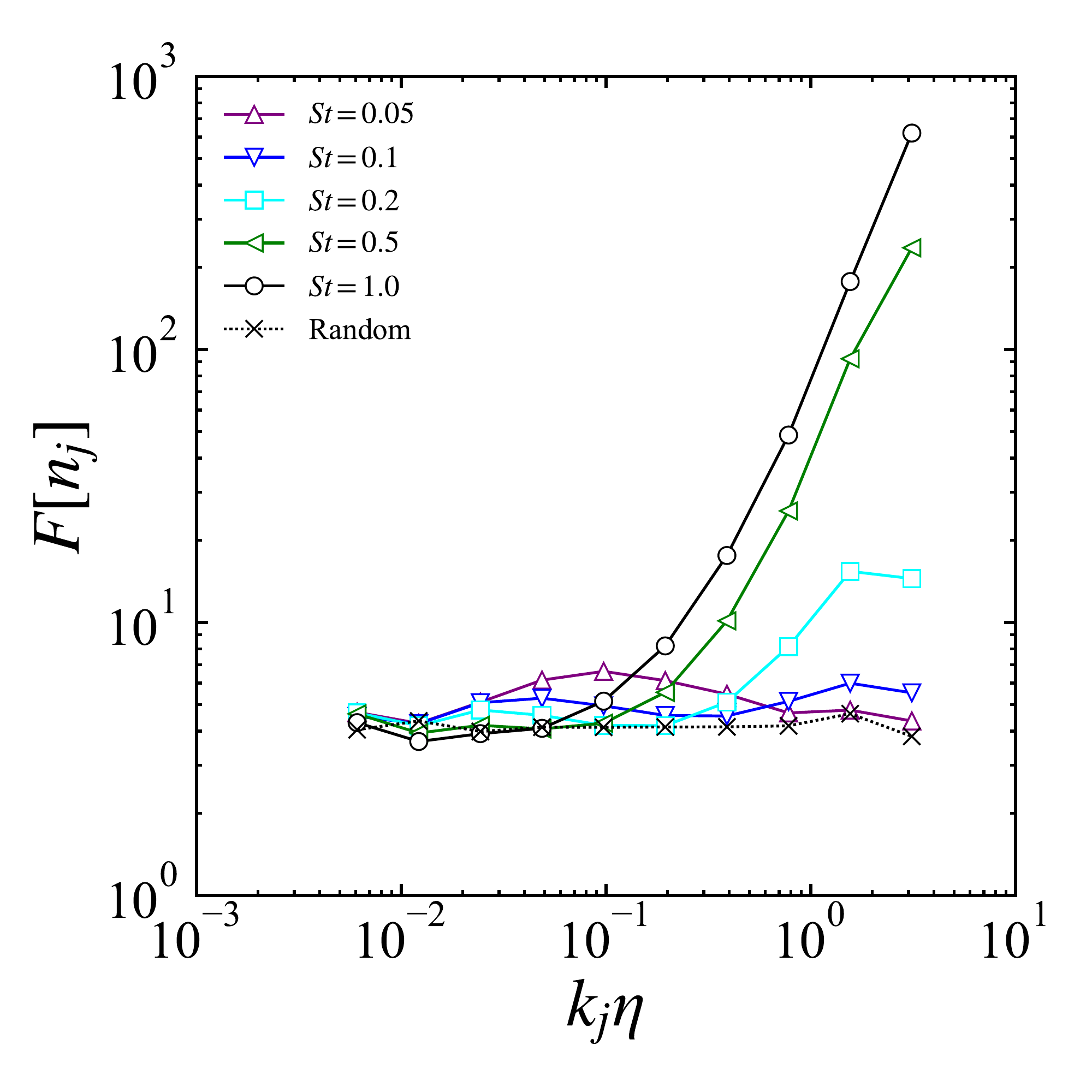} & 
        \includegraphics[width=8cm]{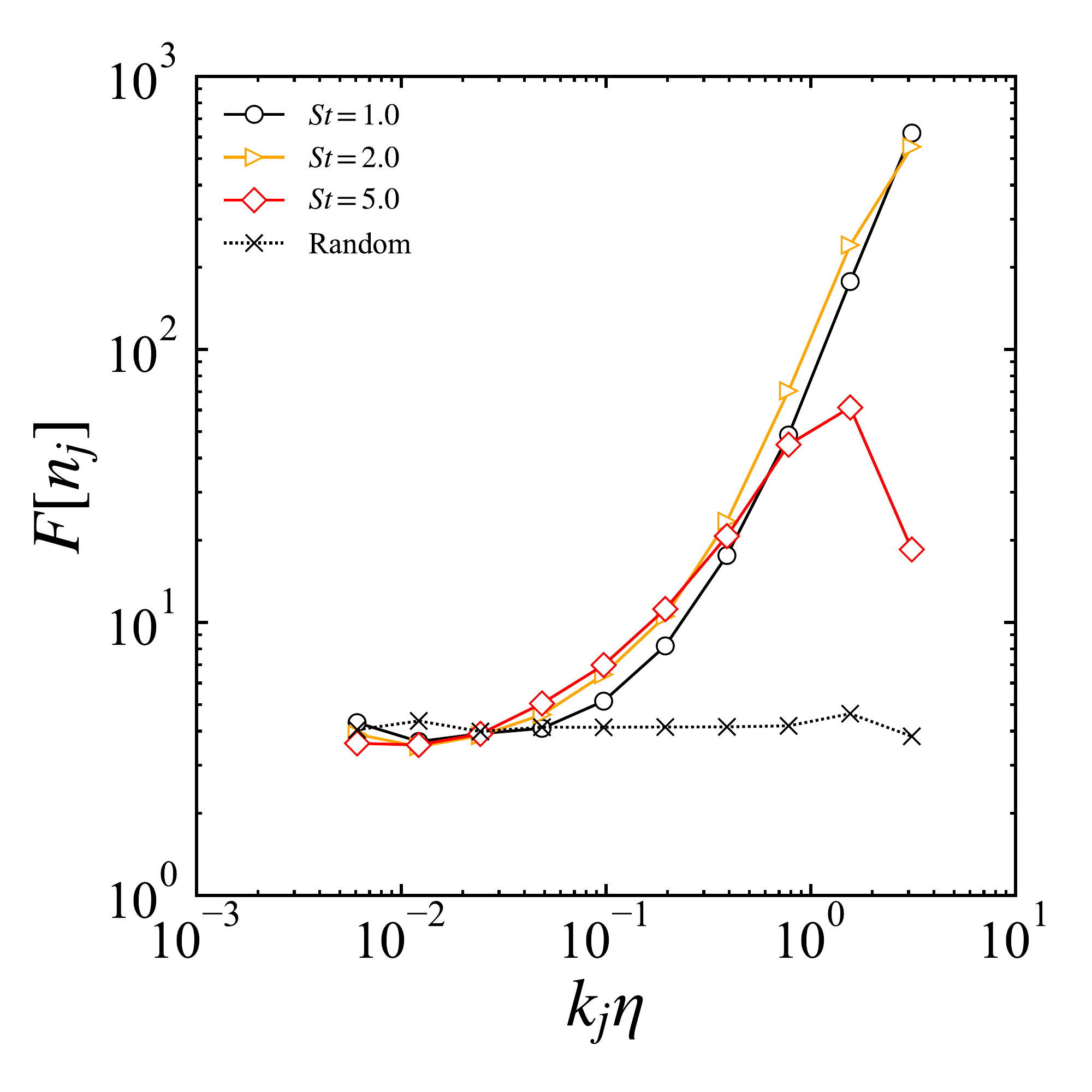}
    \end{tabular}
    \caption{Scale-dependent flatness $F[n_j]$ at $Re_\lambda=204$ for (a) $St\le 1.0$ and (b) $St\ge 1.0$.}
    \label{fig:Fj_St}
\end{figure*}

The Stokes number dependence of $F[n_j]$ and $S[n_j]$ is assessed.
Figure~\ref{fig:Fj_St} shows the scale-dependent flatness $F[n_j]$ for different Stokes numbers.
For the case of $0.5 \le St \le 2.0$, $F[n_j]$ increases as the scale becomes smaller, showing that 
intermittency of clustering is significant in small scales.
For $St=5.0$, $F[n_j]$ decreases at the smallest scale; i.e., 
clusters are less intermittently distributed at the smallest scale. 
This observation for $St=5.0$ is attributed to weak sensitivity of the particles to small eddies. 
The most interesting point in this result is that, for $St \le 0.2$, the flatness $F[n_j]$ at intermediate scales ($0.02 \lesssim k_j \eta  \lesssim 0.4$) increases as the Stokes number decreases. This result is in contradiction to our intuition that the inertial particle distribution becomes 
close to a random distribution as the Stokes number decreases.

\begin{figure*}
    \centering
    \begin{tabular}{ll}
        (a) & (b) \\
        \includegraphics[width=8cm]{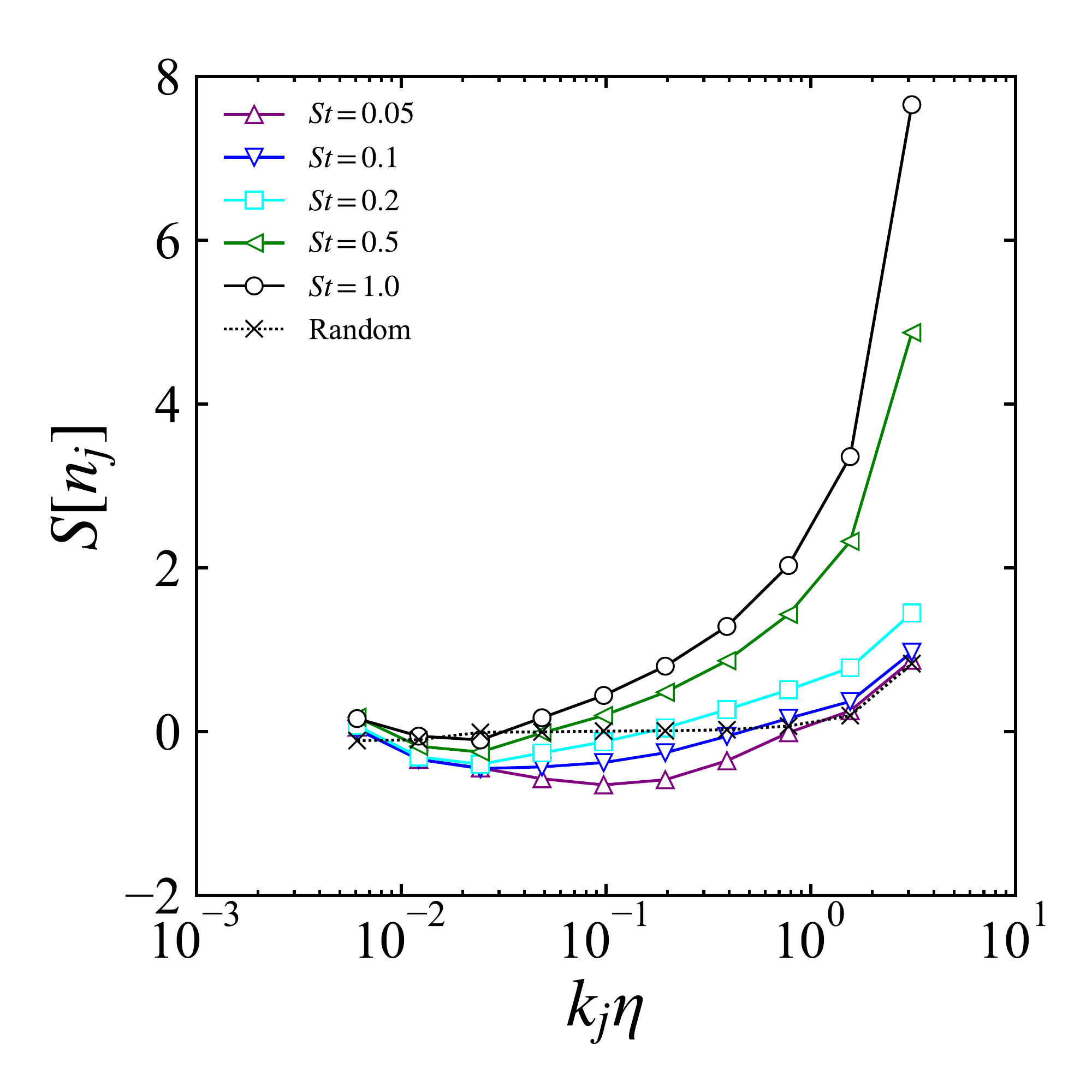} & 
        \includegraphics[width=8cm]{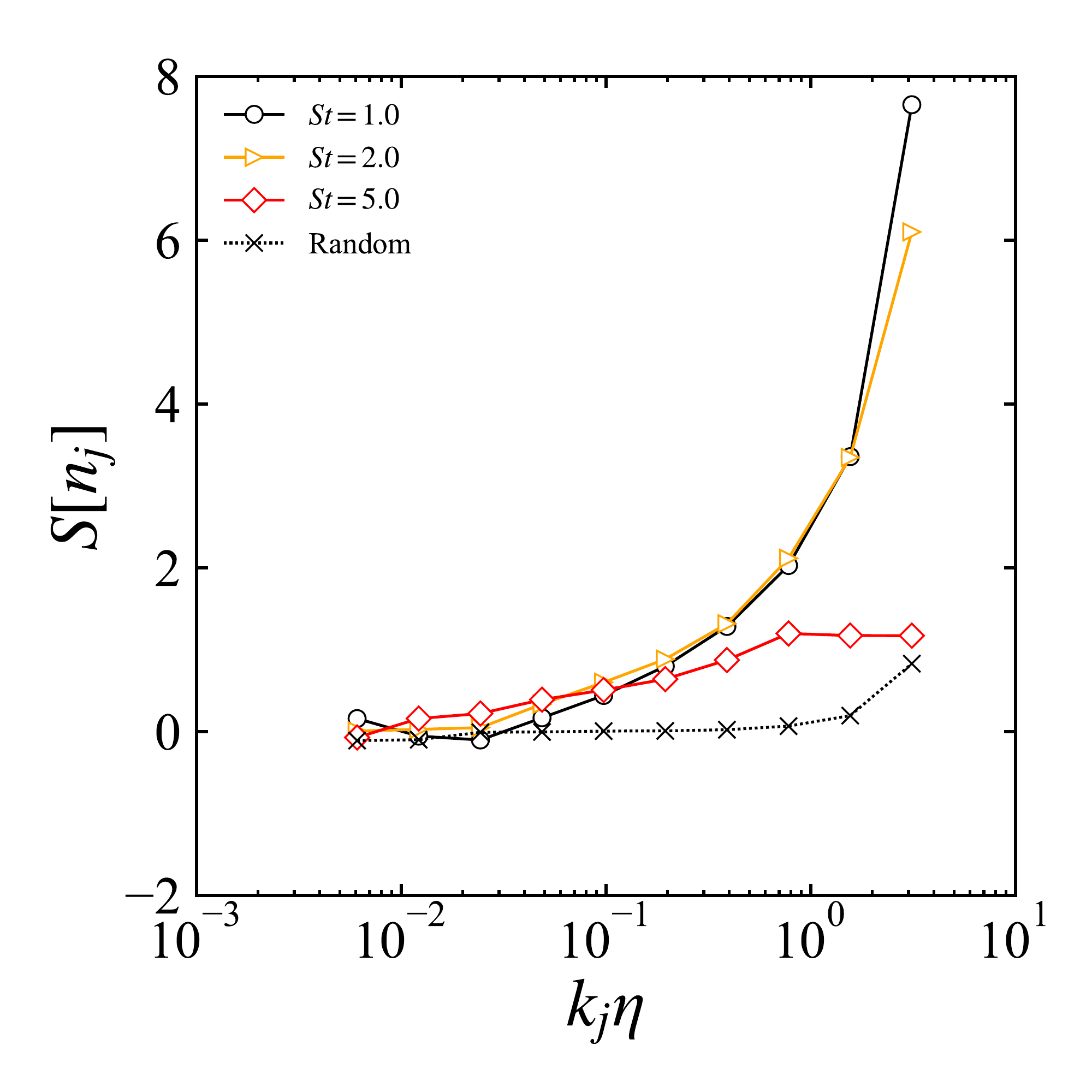}
    \end{tabular}
    \caption{Scale-dependent skewness $S[n_j]$ at $Re_\lambda=204$ for (a) $St\le 1.0$ and (b) $St\ge 1.0$.}
    \label{fig:Sj_St}
\end{figure*}

Figure~\ref{fig:Sj_St} shows the scale-dependent skewness $S[n_j]$ for different Stokes numbers.
For the cases of $0.5 \le St \le 2.0$, $S[n_j]$ increases as the scale becomes smaller, and, for $St=5.0$, $S[n_j]$ saturates for $k_j \eta  \gtrsim 0.8$. 
For $St \le 0.2$, the skewness $S[n_j]$ shows negative values at intermediate scales ($0.02 \lesssim k_j \eta  \lesssim 0.4$).
For the intermediate scales we observe that for $St \le 0.2$ the skewness has locally a concave shape, corresponding in the flatness to a locally convex shape. The local minima of skewness values and the local maxima of the flatness occur at similar scales.
These results suggest that void regions are more pronounced rather than clusters for $St \le 0.2$ while clusters are more pronounced for $St \ge 0.5$.

\subsection{Cluster-pronounced and void-pronounced structures}

\begin{figure*}[tb!]
    \centering
    \begin{tabular}{ll}
        (a) & (b) \\
        \includegraphics[width=8cm]{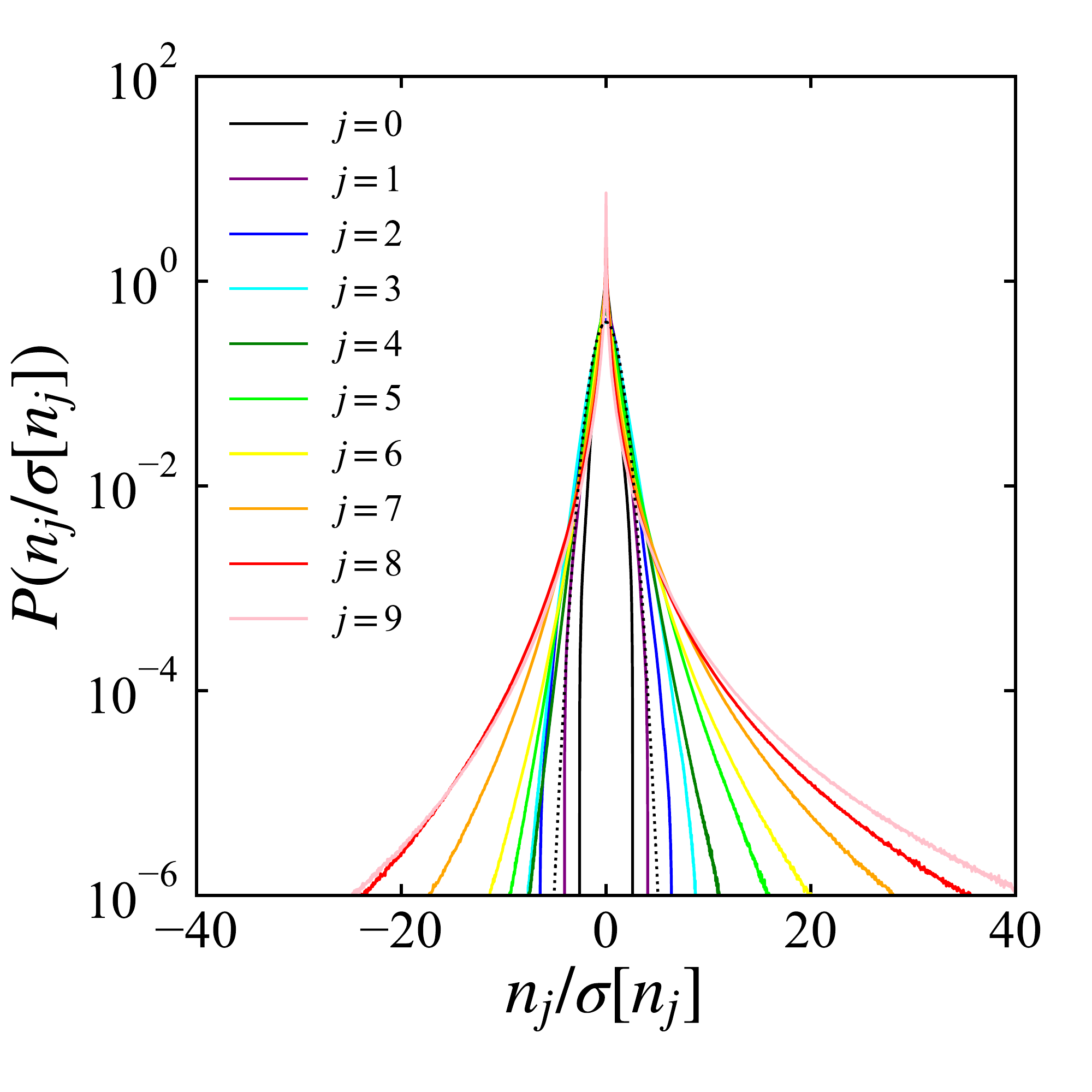}  & 
        \includegraphics[width=8cm]{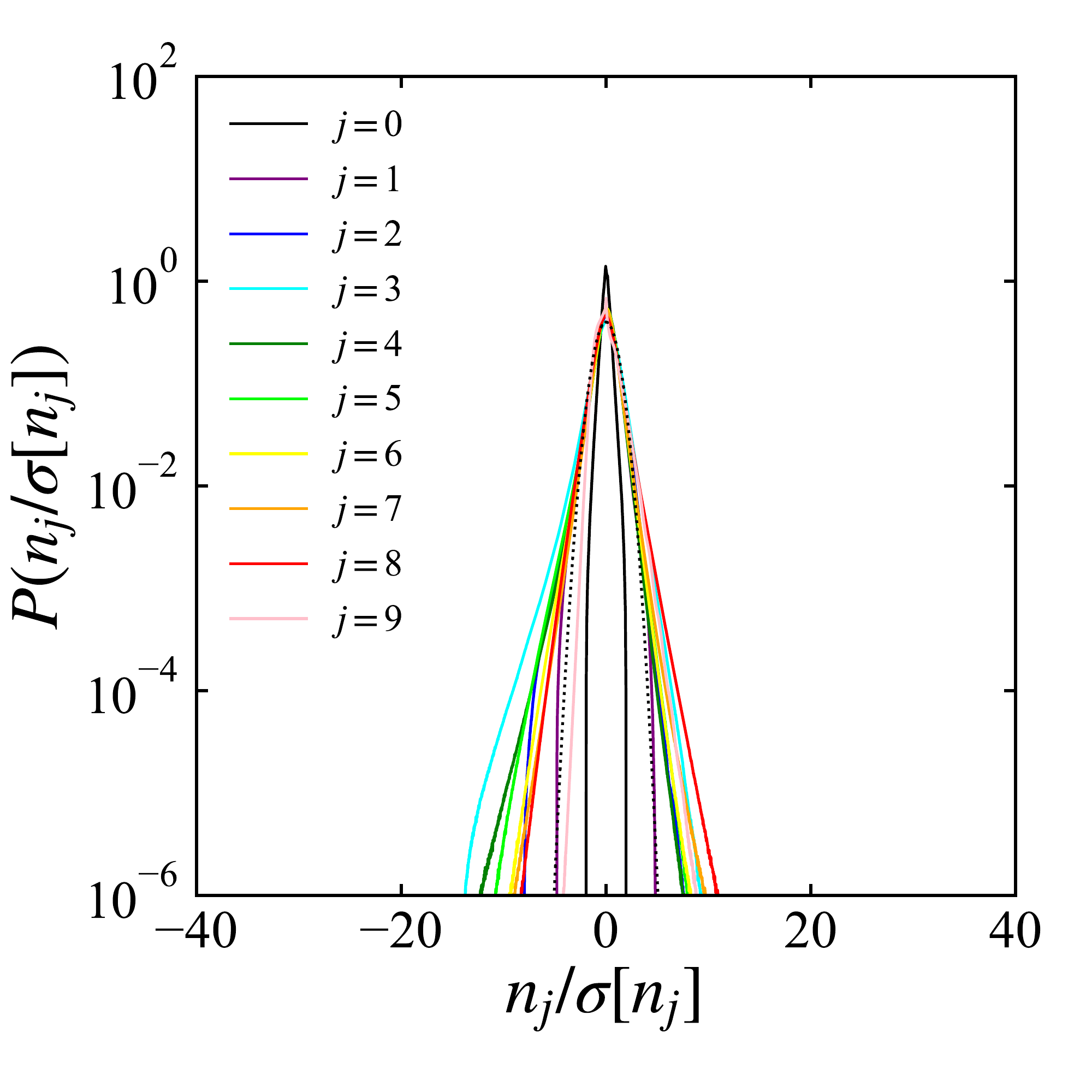}
    \end{tabular}
    \caption{PDF of the normalized scale-dependent particle number density $n_j/\sigma[n_j]$ for (a) $St=1.0$ and (b) $St=0.05$ at $Re_\lambda=204$. The dotted lines are the Gaussian distribution $\mathcal N(0,1)$.}
    \label{fig:PDFnj}
\end{figure*}

\begin{figure}[tb!]
    \centering
    \includegraphics[width=12cm]{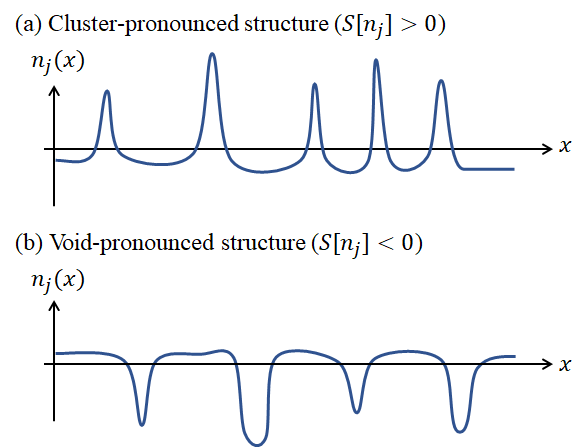}
    \caption{Schematic figures of (a) cluster- and (b) void-pronounced structures for $n_j$.}
    \label{fig:cluster_void_nj}
\end{figure}

\begin{figure*}
    \centering
    \begin{tabular}{ll}
        \includegraphics[height=5.6cm]{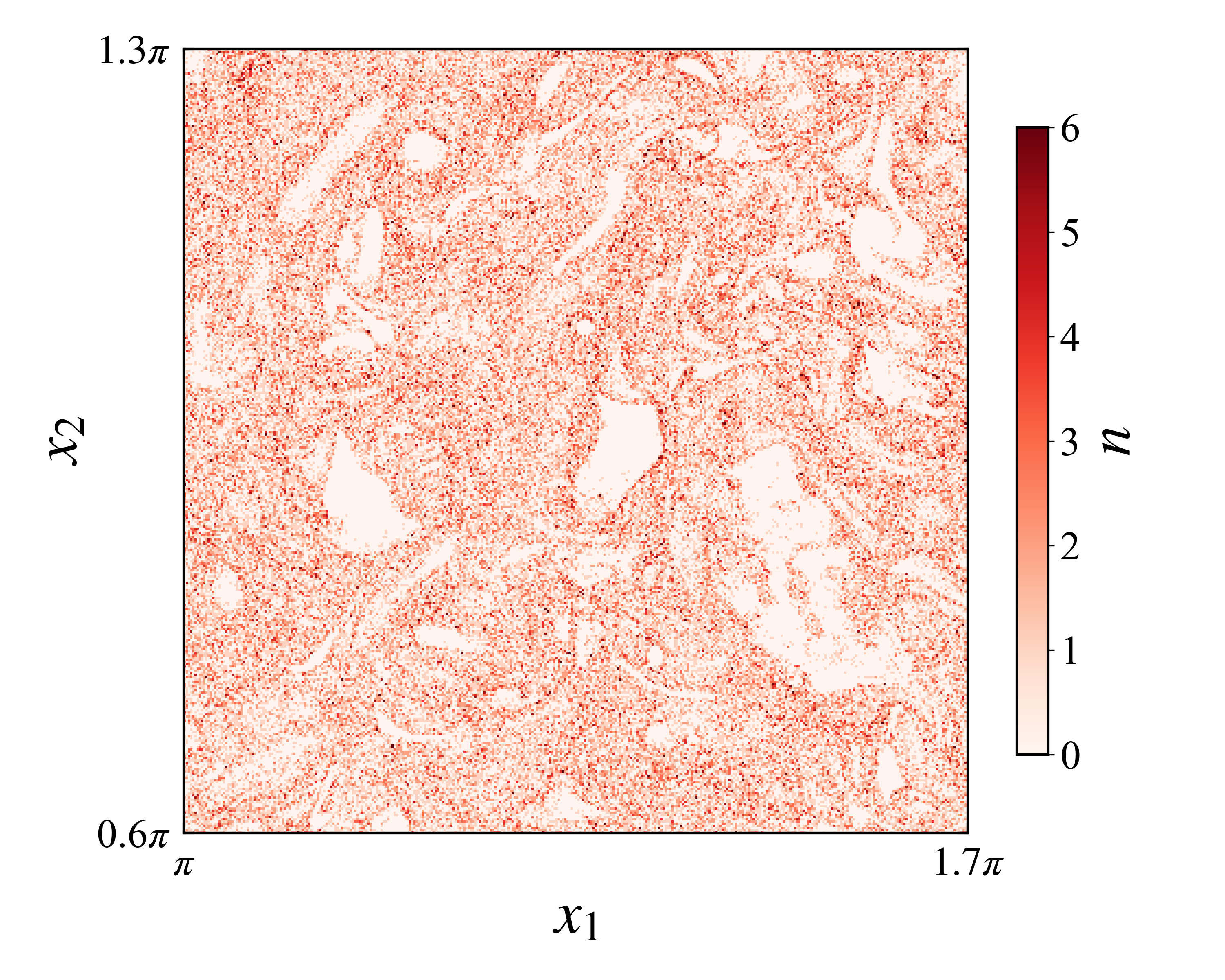} & \includegraphics[height=5.6cm]{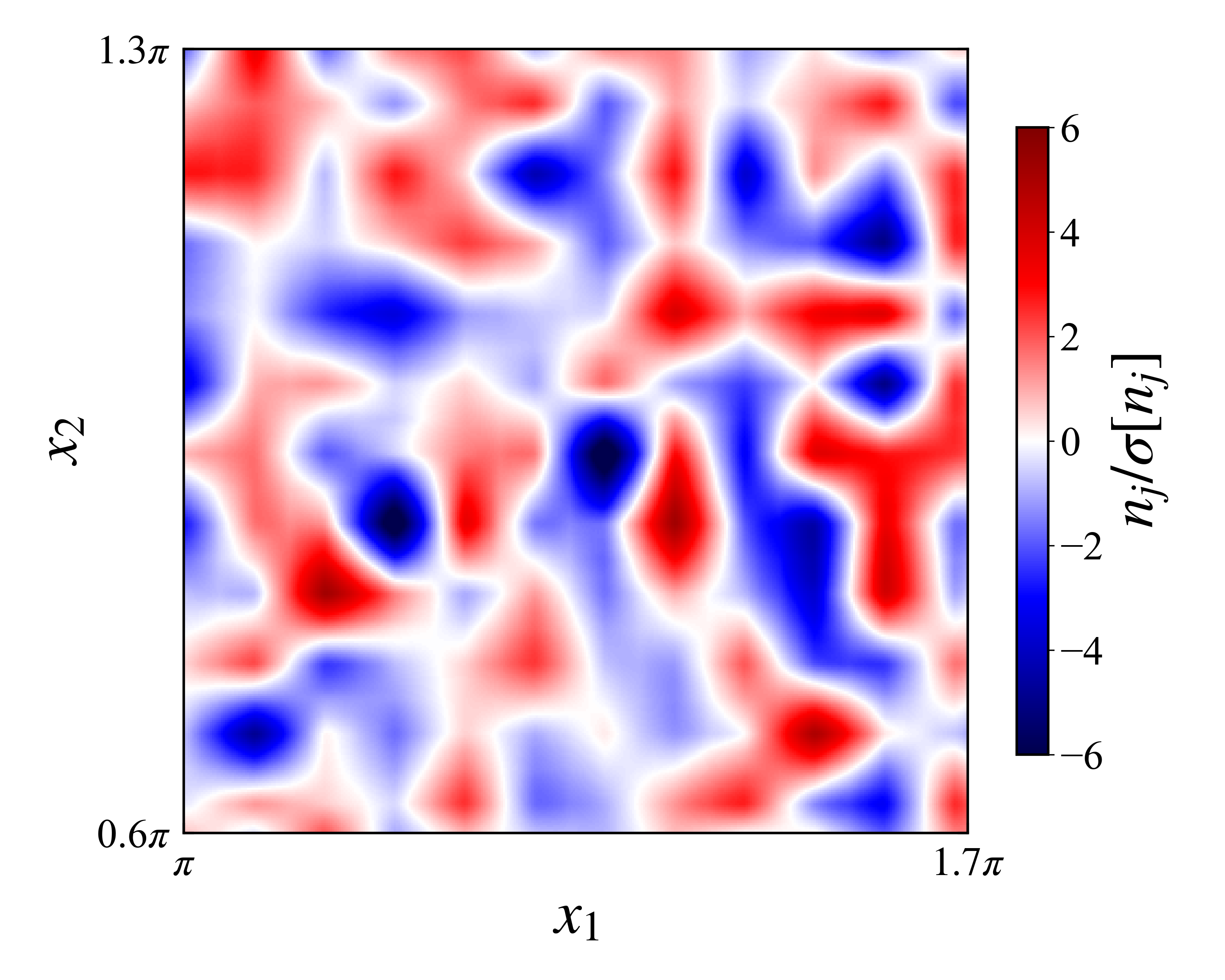} \vspace{-56mm} \\
        (a) & (b) \vspace{44mm}
    \end{tabular}
    \caption{Magnified spatial distributions of (a) total number density $n({\bm x})$ and (b) scale contributions $n_j({\bm x})$ at $j=4$ for $St=0.05$ in the same $x_1$-$x_2$ cross section as Figure~\ref{fig:density2d}.}
    \label{fig:density2d_zoom}
\end{figure*}

To clarify the difference of cluster-pronounced and void-pronounced structures, 
the PDFs of $n_j({\bm x})$ normalized by the standard deviation $\sigma[n_j]$ for $St=1.0$ and $St=0.05$ are shown in Fig.~\ref{fig:PDFnj}. 
For $St=1.0$, skewness $S[n_j]$ is positive for $k_j \eta \gtrsim 0.05$ ($j=4,\cdots,9$) as shown in Fig.~\ref{fig:Sj_St}(a). The PDF in Fig.~\ref{fig:PDFnj}(a) has a heavier tail on the positive side for each $j$ ($j=4,\cdots,9$).
In contrast, for $St=0.05$, the PDF in Fig.~\ref{fig:PDFnj}(b) has a heavier tail on the negative side for each $j$ ($j=2, \cdots, 6$), where 
$S[n_j]$ is negative (i.e., $0.02 \lesssim k_j \eta  \lesssim 0.4$) as shown in Fig.~\ref{fig:Sj_St}(a).
These trends in the PDFs imply that $n_j({\bm x})$ has higher probability of large positive values when $S[n_j]>0$, while $n_j({\bm x})$ has higher probability of large negative values when $S[n_j]<0$.
Thus, the spatial distribution of $n_j({\bm x})$ is expected to behave like in the schematic figures in Fig.~\ref{fig:cluster_void_nj}.
To clarify whether negative skewness is a sign of void-pronounced structures, we verify 
the relationship between the large negative values of $n_j({\bm x})$ and void regions.
Figures~\ref{fig:density2d_zoom}(a) and \ref{fig:density2d_zoom}(b) respectively show magnified views of Figs.~\ref{fig:density2d}(b) and \ref{fig:density2d}(f), which are the total number density $n({\bm x})$ and the scale contribution for $j=4$, corresponding to $k_4 \eta = 9.7\times10^{-2}$. 
Note that the scale index $j=4$ corresponds to the scale at which the skewness value is minimum at this Stokes number, $St=0.05$.
Figure~\ref{fig:density2d_zoom} shows that the location of large negative values in $n_j ({\bm x})$ corresponds to void regions in $n({\bm x})$. We can therefore conclude that for $St \le 0.2$ negative skewness values are indicators for void-pronounced structures.

For $St \le 0.2$, the scales of local minima of the negative skewness in Fig.~\ref{fig:Sj_St}(a) almost correspond to the scales of local maxima of the flatness in Fig.~\ref{fig:Fj_St}(a). 
This suggests that the flatness at intermediate scales for $St \le 0.2$ is attributed to the intermittent distribution of void regions.
Thus the intermittent void distributions play an important role for inertial particle clustering for $St \le 0.2$. 
%
Negative skewness values can also be observed in Fig.~\ref{fig:SjFj_Re}: for $Re_\lambda \ge 328$, $S[n_j]$ shows negative values at large scales $k_j\eta \lesssim 0.02$. It is conjectured that void regions are pronounced at large scales, while clusters are pronounced at small scales.
This result could be connected to ``cloud voids'' reported by Karpi\'nska {\it et al.}~\cite{Karp2019}. They observed many void regions with the diameter of up to $12$ cm during mountain observations and explained that the phenomenon is caused by the inertial motion of cloud particles.


\section{Conclusion}\label{sec5}
We have studied scale-dependent statistics of the particle distribution to get insight into the nonuniform distribution of inertial particles, i.e., clusters and voids, in isotropic turbulence. 
To this end orthogonal wavelet analyses have been applied to 
particle data obtained by performing three-dimensional direct numerical simulation of particle-laden homogeneous isotropic turbulence
at high Reynolds number ($Re_\lambda \gtrsim 200$) using up to $10^9$ particles.  
%
The number density fields $n({\bm x},t)$ 
are obtained by the histogram method using equidistant bins, 
and are then decomposed into scale-dependent contributions $n_j({\bm x},t)$ at scale $2^{-j}$ 
using orthogonal wavelet filtering.
Scale-dependent skewness and flatness values 
have been investigated and the influence of the Reynolds and Stokes number
has been assessed. 
The following conclusions can be drawn.


For $St=1.0$ the influence of Reynolds number $Re_\lambda$ was assessed. 
We found that the scale-dependent flatness $F[n_j]$ increases slightly as $Re_\lambda$ increases at scales larger than the Kolmogorov scale, 
while the $Re_\lambda$ dependence of the scale-dependent skewness $S[n_j]$ is negligibly small. 

We observed that the influence of the Stokes number $St$ on $F[n_j]$ and $S[n_j]$ is more significant compared to the influence of the considered $Re_\lambda$.
For $0.5 \le St \le 2.0$, both the scale-dependent skewness and flatness values
become larger, when the scale decreases. 
This suggests intermittent clustering at small scales. The intermittency is reflected by the increasing flatness values, while the clustering can be explained by the increasing skewness values.
%
For $St>1.0$, we observe that the flatness at the smallest scale becomes smaller as $St$ increases, which means that the particle number density becomes less intermittent.
%
We also found that for small Stokes numbers, $St \le 0.2$, the skewness $S[n_j]$ exhibits negative values at intermediate scales, %
 i.e. for scales larger than the Kolmogorov scale and smaller than the integral scale of the flow,
and the flatness $F[n_j]$ at the intermediate scales increases as $St$ decreases. 
We have shown that negative values of $S[n_j]$ imply higher probability of large negative values of $n_j$. Our visualizations show that these large negative values of $n_j$ can be attributed to void regions of the particle number density. 
Hence we can conclude that void regions at the intermediate scales are pronounced and intermittently distributed for $St \le 0.2$.
We conjecture
that intermittent void distributions play an important role for inertial particle clustering for $St \le 0.2$.
%
Our results for higher Reynolds numbers, i.e., for $Re_\lambda = 328$ and $531$, confirm that negative values of the skewness $S[n_j]$ are likewise observed at large scales.
This suggests that void regions are pronounced at large scales, while clusters are pronounced at small scales.

%
The dynamics of scale-dependent cluster and void formation is still an open issue and its clarification is of importance for modeling. The divergence of the particle velocity plays hereby a key role, as recently shown in Ref. \cite{TK2}. Analyzing the dynamics of the scale-dependent divergence is an interesting perspective for future work.



\vspace*{1cm}
\begin{acknowledgments}
KM acknowledges financial support from Grant-in-Aid for Young Scientists (B) No. JP17K14598 and Scientific Research (C) No. JP20K04298 from the Japan Society for the Promotion of Science (JSPS). He is also grateful to I2M for kind hospitality during his stay at Aix-Marseille Universit\'e.
KY acknowledges financial support from Grant-in-Aid for Scientific Research (S) No. JP16H06339 from the JSPS, and thanks I2M for kind hospitality during his stay at Aix-Marseille Universit\'e. 
KS acknowledges kind hospitality during his visit at JAMSTEC and financial support.
The direct numerical simulations presented here were conducted on the Earth Simulator supercomputer system operated by JAMSTEC.
\end{acknowledgments}

\appendix
\section{Grid number and particle number dependence}\label{gnpn}

\begin{figure*}[p!]
    \centering
    \begin{tabular}{ll}
        \includegraphics[width=7.2cm]{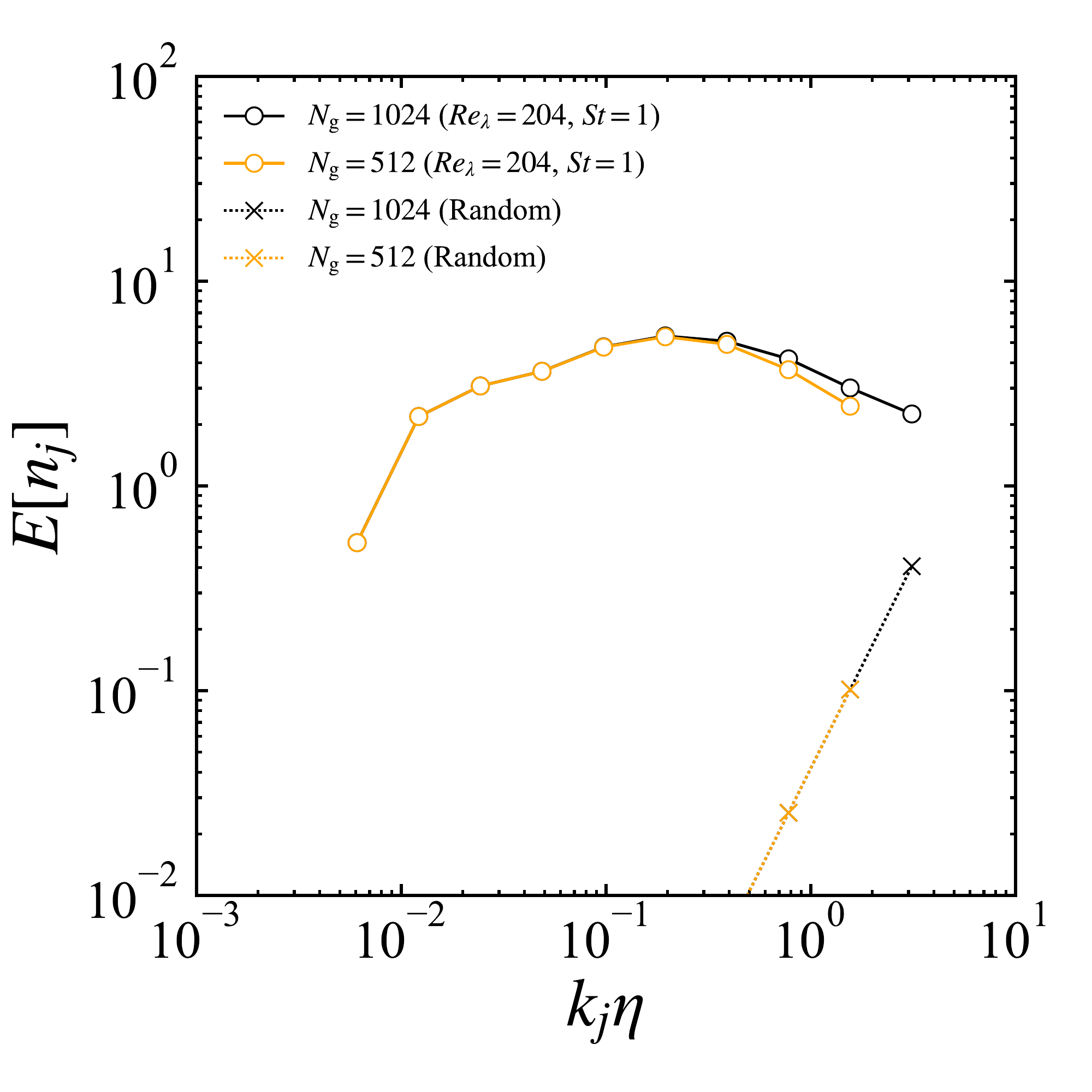} & 
        \includegraphics[width=7.2cm]{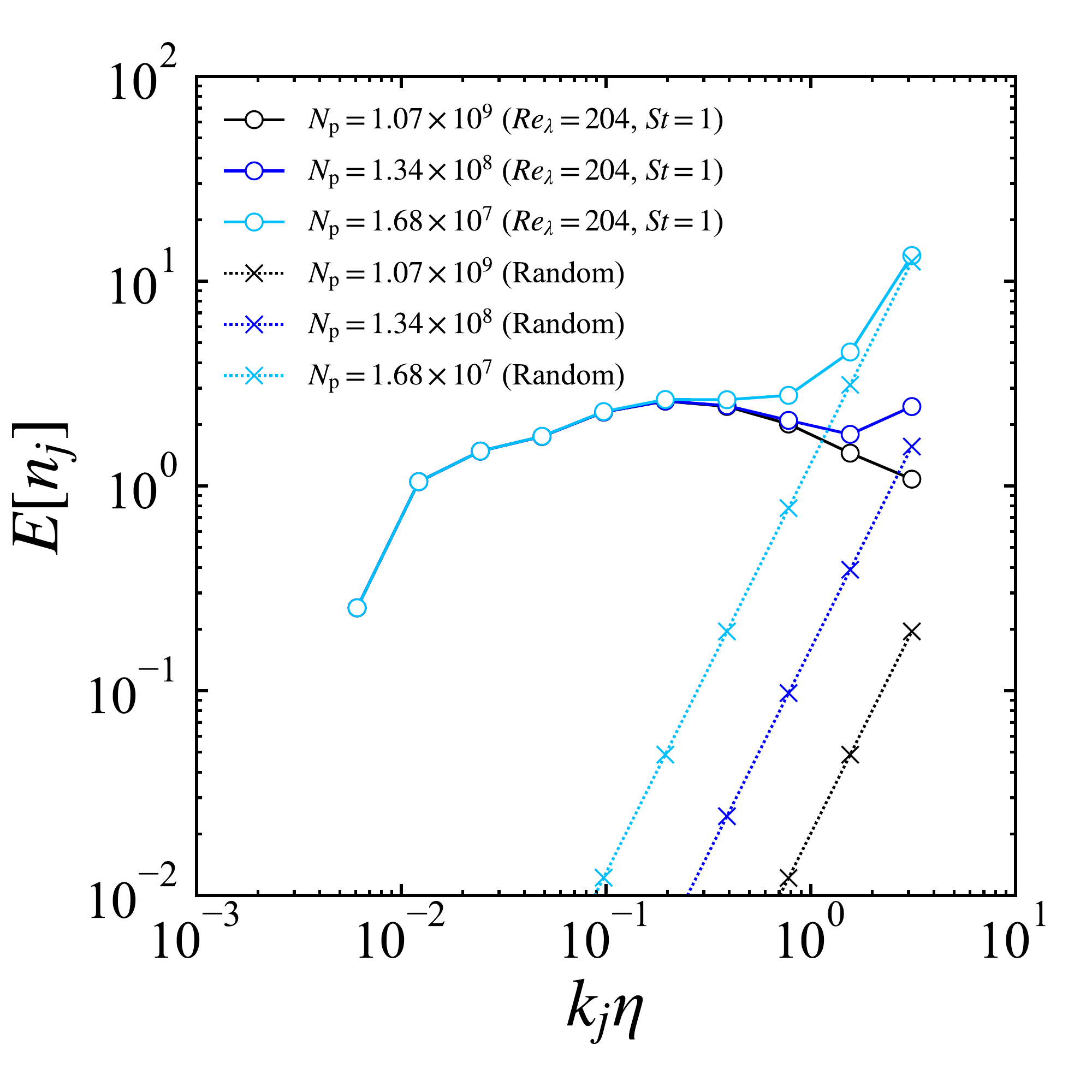} \vspace{-72mm} \\
        (a) & (b) \vspace{60mm} \\
        \includegraphics[width=7.2cm]{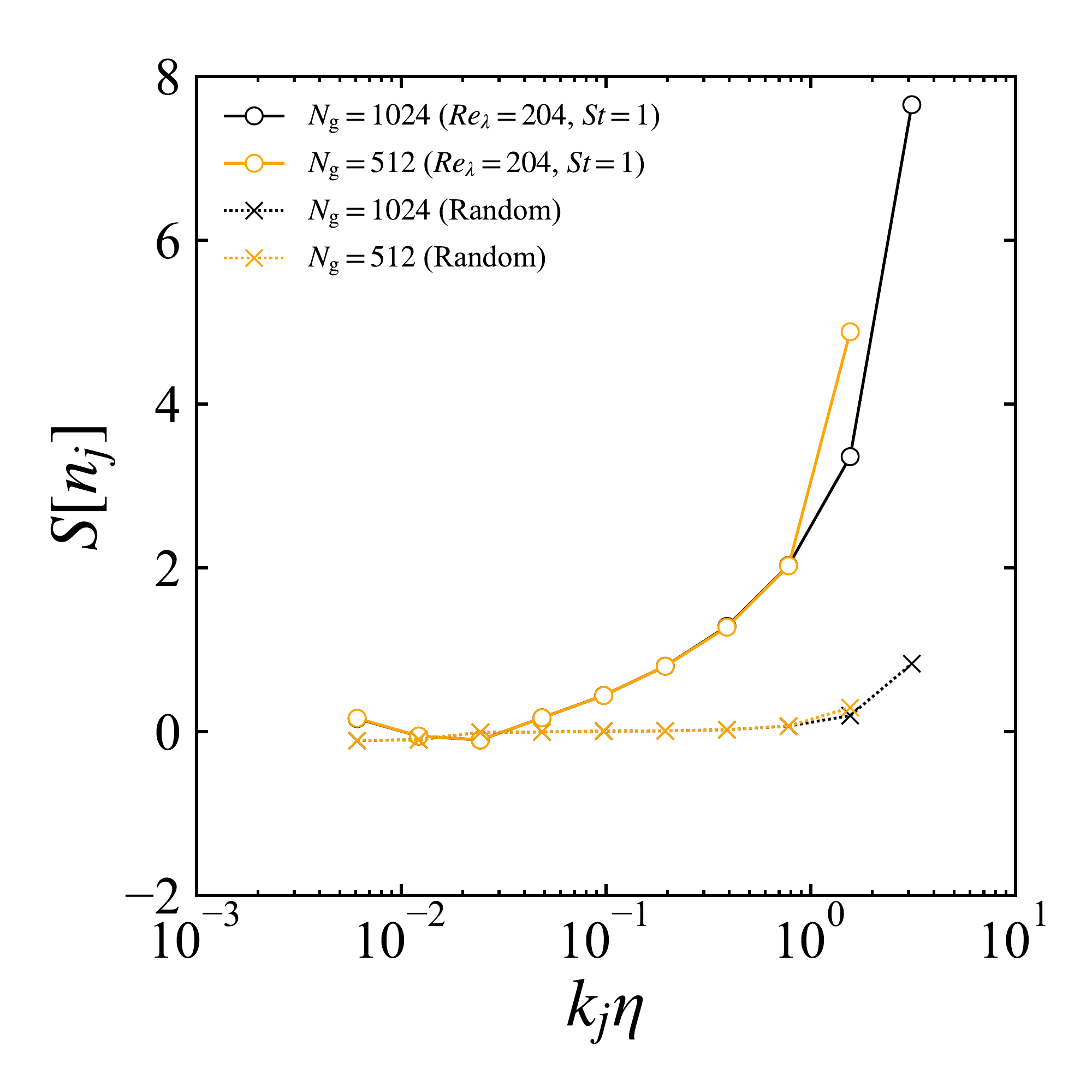} & 
        \includegraphics[width=7.2cm]{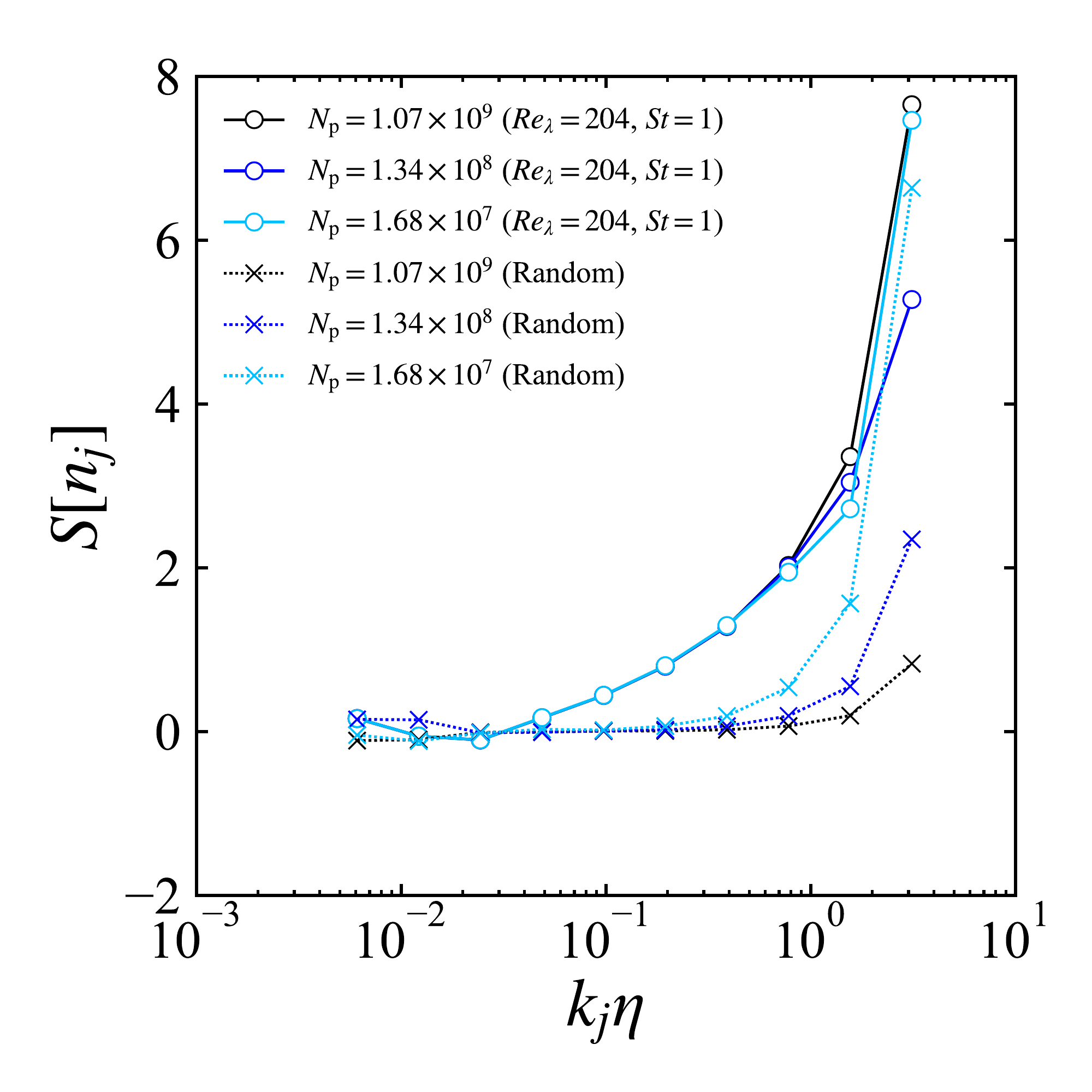} \vspace{-72mm} \\
        (c) & (d) \vspace{60mm} \\
        \includegraphics[width=7.2cm]{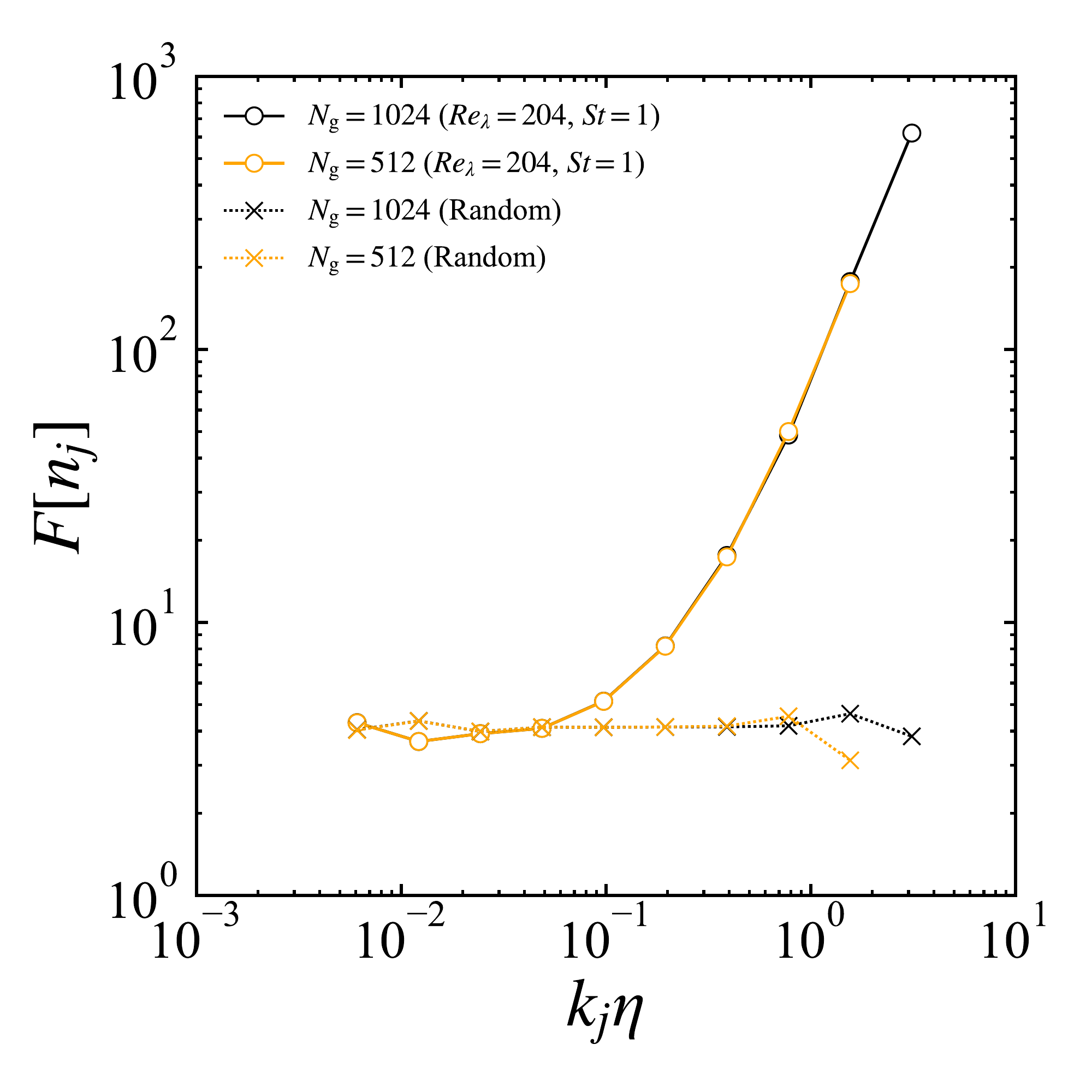} & 
        \includegraphics[width=7.2cm]{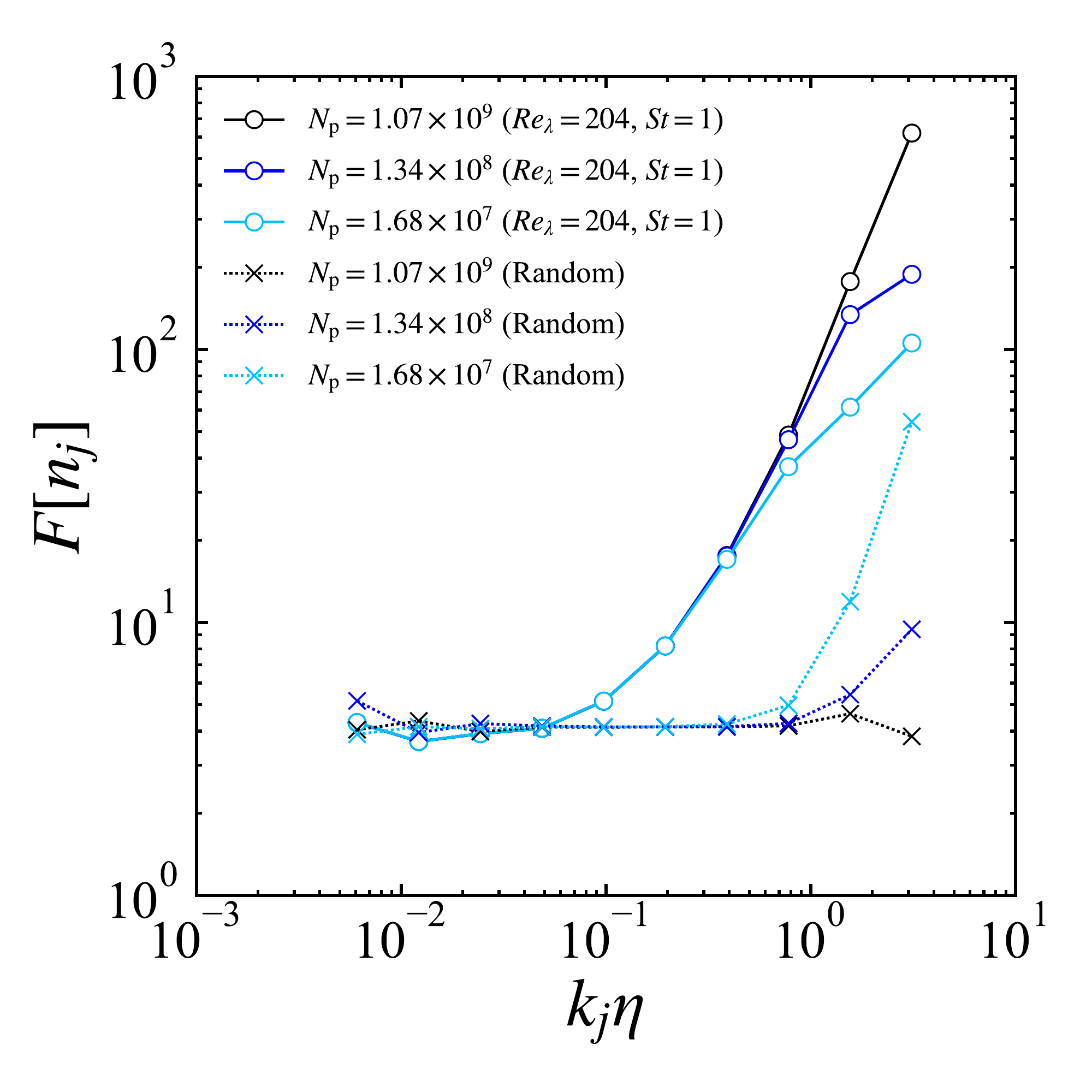} \vspace{-72mm} \\
        (e) & (f) \vspace{60mm}
    \end{tabular}
    \caption{Wavelet spectra $E[n_j]$ (a, b), scale-dependent skewness $S[n_j]$ (c, d) and scale-dependent flatness $F[n_j]$ (e, f) at $Re_\lambda=204$ and $St=1.0$ for (a, c, e) $N_g=512$ and $1024$ at fixed $N_p (=1.07\times10^9)$ and for (b, d, f) $N_p=1.68\times10^7$, $1.34\times10^8$ and $1.07\times10^9$ at fixed $N_g (=1024)$. }
    \label{fig:EjSjFj_NgNp}
\end{figure*}

The influence of the numerical parameters, i.e., the number of grid points $N_g$ and the number of particles $N_p$ in simulations, can be
crucial when performing statistical analyses, especially for higher order statistics. We check the influence of these parameters on the energy spectra $E[n_j]$, scale-dependent skewness $S[n_j]$ and flatness $F[n_j]$ of particle number density fields $n({\bm x})$ in the DNS for $St=1.0$ and $Re_\lambda =204$.
In addition, we compare them with randomly distributed particles.
 Figure~\ref{fig:EjSjFj_NgNp} quantifies the impact of $N_g$ and $N_p$ on the scale-dependent statistics, $E[n_j]$, $S[n_j]$ and $F[n_j]$, plotted as a function of $k_j\eta$.
%
%
Figure~\ref{fig:EjSjFj_NgNp}(a) illustrates that doubling $N_g$ from 512 to 1024 has a small influence on the energy spectrum of inertial particles at small scales due to the difference of the filter size for the histogram method in Eq.~(\ref{eq:particlenumberfield}), while the doubling does not impact the spectrum for randomly distributed particles which exhibits a $k^2$ behavior~\cite{Bassenne2017,SaitoGotoh2018}. 
Changing the number of particles $N_p$, while keeping the grid size fixed ($N_g = 1024$), shows some impact on the spectra at small scales in Fig.~\ref{fig:EjSjFj_NgNp}(b). 
Figure~\ref{fig:EjSjFj_NgNp}(b)
nicely quantifies the influence of the particle noise. 
The spectra for randomly distributed particles are shown for comparison. 
The latter are proportional to $k_j^2/N_p$
and we can see that at small scales the spectra of the particle fields are polluted with noise contributions implying a $k^2$ behavior. 
Figure~\ref{fig:EjSjFj_NgNp}(c) shows that the number of grid points $N_g$ has 
small impact on $S[n_j]$ only at the smallest scale of each $N_g$.
The influence of $N_g$ becomes even weaker for $F[n_j]$, as shown in Fig.~\ref{fig:EjSjFj_NgNp}(e). For inertial particles both quantities grow significantly with decreasing scale, i.e., increasing $k_j \eta$.
For the random cases, we observe that for each $N_g$, $S[n_j]$ increases only weakly with decreasing scale and $F[n_j]$ even remains almost constant.
However, for inertial particles the number of particles $N_p$ has some impact on both $S[n_j]$ and $F[n_j]$. This is also the case for the random particles. 
For inertial particles we find that the growth of $S[n_j]$ and $F[n_j]$ with $k_j \eta$ becomes more pronounced when increasing the number of particles from $N_p = 1.68 \times 10^7$ to $N_p = 1.07 \times 10^9$, as observed in Figs.~\ref{fig:EjSjFj_NgNp}(d) and \ref{fig:EjSjFj_NgNp}(f).
For the random case this trend is inverted: increasing $N_p$ yields more stable statistical estimators and thus the growth of $S[n_j]$ and $F[n_j]$ with $k_j \eta$ is reduced. 
The current results suggest that in the case of random particles $S[n_j]$ and $F[n_j]$  increase for $ 2^{-3(j+1)} N_p \lesssim 1$. 
As for randomly distributed particles, void and cluster regions are absent, the skewness values should vanish and the flatness values should remain constant with scale. In other words, deviation of the skewness and flatness values for the random case is caused by statistical sampling, i.e., the finite numbers of particles $N_p$.
The above observations illustrate the importance to use a sufficiently large number of particles to get statistically converged results and to observe skewness and flatness values independent of $N_p$.
The increasing values of $S[n_j]$ and $F[n_j]$ with $k_j\eta$, i.e., for decreasing scale, in the random cases can thus be used to determine whether $N_p$ is sufficiently large or not.


\begin{thebibliography}{99}
\bibitem{Shaw} R. A. Shaw, Particle-turbulence interactions in atmospheric clouds, Annu. Rev. Fluid Mech. {\bf 35}, 183 (2003).
\bibitem{Ireland2016} P. J. Ireland, A. D. Bragg, and L. R. Collins, The effect of Reynolds number on inertial particle dynamics in isotropic turbulence. Part 1. Simulations without gravitational effects, 
J. Fluid Mech. {\bf 796}, 617 (2016).
\bibitem{Matsuda2014} K. Matsuda, R. Onishi, M. Hirahara, R. Kurose, K. Takahashi, and S. Komori, Influence of microscale turbulent droplet clustering on radar cloud observations, J. Atmos. Sci. {\bf 71}, 3569 (2014).
\bibitem{MO2019} K. Matsuda and R. Onishi, Turbulent enhancement of radar reflectivity factor for polydisperse cloud droplets, Atmos. Chem. Phys. {\bf 19}, 1785 (2019).
\bibitem{Toschi} F. Toschi and E. Bodenschatz, Lagrangian properties of particles in turbulence, Annu. Rev. Fluid Mech. {\bf 41}, 375 (2009).
\bibitem{Monchaux2012} R. Monchaux, M. Bourgoin, and A. Cartellier, Analyzing preferential concentration and clustering of inertial particles in turbulence, Int. J. Multiphase Flow {\bf 40}, 1 (2012).
\bibitem{Maxey} M. R. Maxey, The gravitational settling of aerosol particles in homogeneous turbulence and random flow fields, J. Fluid Mech. {\bf 174}, 441 (1987).
\bibitem{Squires} K. D. Squires and J. K. Eaton, Measurements of particle dispersion obtained from direct numerical simulations of isotropic turbulence, J. Fluid Mech. {\bf 226}, 1 (1991). 
\bibitem{Sundaram_Collins1997} S. Sundaram and L. R. Collins,  Collision statistics in an isotropic particle-laden turbulent suspension. Part 1. Direct numerical simulations, J. Fluid Mech. {\bf 335}, 75 (1997).
\bibitem{Wang2000} L.-P. Wang, A. Wexler, and Y. Zhou, Statistical mechanical description and modelling of turbulent collision of inertial particles, J. Fluid Mech. {\bf 415}, 117 (2000).
\bibitem{Boffetta2004} G. Boffetta, F. De Lillo, and A. Gamba, Large scale inhomogeneity of inertial particles in turbulent flows, Phys. Fluids {\bf 16}, L20 (2004).
\bibitem{YG2007} H. Yoshimoto and S. Goto, Self-similar clustering of inertial particles in homogeneous turbulence, J. Fluid Mech. {\bf 577}, 275 (2007).
\bibitem{CV2009} S. W. Coleman and J. C. Vassilicos, A unified sweep-stick mechanism to explain particle clustering in two- and three-dimensional homogeneous, isotropic turbulence, Phys. Fluids {\bf 21}, 113301 (2009).
\bibitem{GV2006} S. Goto and J. C. Vassilicos, Self-similar clustering of inertial particles and zero-acceleration points in fully developed two-dimensional turbulence, Phys. Fluids {\bf 18},  115103 (2006).
\bibitem{CGV2006} L. Chen, S. Goto, and J. C. Vassilicos, Turbulent clustering of stagnation points and inertial particles, J. Fluid Mech. {\bf 553}, 143 (2006).
\bibitem{GV2008} S. Goto and J. C. Vassilicos, Sweep-Stick Mechanism of Heavy Particle Clustering in Fluid Turbulence, Phys. Rev. Lett. {\bf 100}, 054503 (2008).
\bibitem{Monchaux2010} R. Monchaux, M. Bourgoin, and A. Cartellier, Preferential concentration of heavy particles: A Vorono\"i analysis, Phys. Fluids {\bf 22}, 103304 (2010).
\bibitem{Bec2007} J. Bec, L. Biferale, M. Cencini, A. Lanotte,  S. Musacchio, and F. Toschi, Heavy Particle Concentration in Turbulence at Dissipative and Inertial Scales. Phys. Rev. Lett. {\bf 98}, 084502 (2007).
\bibitem{Saw2008} E. W. Saw, R. A. Shaw, S. Ayyalasomayajula, P. Y. Chuang, and \'A. Gylfason, Inertial Clustering of Particles in High-Reynolds-Number Turbulence, Phys. Rev. Lett. {\bf 100}, 214501 (2008).
\bibitem{Bragg2015} A. D. Bragg, P. J. Ireland, and L. R. Collins, Mechanisms for the clustering of inertial particles in the inertial range of isotropic turbulence, Phys. Rev. E {\bf 92}, 023029 (2015).
\bibitem{Ariki2018} T. Ariki, K. Yoshida, K. Matsuda, and K. Yoshimatsu, Scale-similar clustering of heavy particles in the inertial range of turbulence, Phys. Rev. E {\bf 97}, 033109 (2018).%
\bibitem{Bassenne2017} M. Bassenne, J. Urzay, K. Schneider, and P. Moin, Extraction of coherent clusters and grid adaptation in particle-laden turbulence using wavelet filters, Phys. Rev. Fluids {\bf 2}, 054301 (2017).
\bibitem{Bassenne2018} M. Bassenne, P. Moin, and J. Urzay, Wavelet multiresolution analysis of particle-laden turbulence, Phys. Rev. Fluids {\bf 3}, 084304 (2018).
\bibitem{FS89} M. Farge and R. Sadourny, Wave-vortex dynamics in rotating shallow water, J. Fluid Mech. {\bf 206}, 433 (1989).
\bibitem{Fa92} M. Farge, Wavelet transforms and their applications to turbulence, Annu. Rev. Fluid Mech. {\bf 24}, 395 (1992).
\bibitem{Mene91} C. Meneveau, Analysis of turbulence in the orthonormal wavelet representation, J.~Fluid Mech. {\bf 232}, 469 (1991).
\bibitem{FSK99} M. Farge, K. Schneider, and N. Kevlahan, Non-Gaussianity and coherent vortex simulation for two-dimensional turbulence using an adaptive orthogonal wavelet basis, Phys. Fluids {\bf 11}, 2187 (1999).
\bibitem{FPS01} M. Farge, G. Pellegrino, and K. Schneider, Coherent Vortex Extraction in 3D Turbulent Flows Using Orthogonal Wavelets, Phys. Rev. Lett. {\bf 87}, 054501 (2001).
\bibitem{OYSFK07} N. Okamoto, K. Yoshimatsu, K. Schneider, M. Farge, and Y.  Kaneda, Coherent vortices in high resolution direct numerical simulation of homogeneous isotropic turbulence: A wavelet viewpoint, Phys. Fluids {\bf 19}, 115109 (2007). 
\bibitem{SFK04} K.~Schneider, M.~Farge, and N.~Kevlahan, {\it Spatial intermittency in two-dimensional turbulence: a wavelet approach, in Woods Hole Mathematics, Perspectives in Mathematics and Physics}, edited by N.~Tongring and R.~C.~Penner (World Scientific, Singapore, 2004), Vol. 34, pp. 302--328.
\bibitem{BLS07} W. J. Bos, L. Liechtenstein, and K. Schneider, Small-scale intermittency in anisotropic turbulence, Phys. Rev. E {\bf 76}, 046310 (2007).
\bibitem{YOSKF09} K. Yoshimatsu, N. Okamoto, K. Schneider, Y. Kaneda, and M. Farge, Intermittency and scale-dependent statistics in fully developed turbulence, Phys. Rev. E {\bf 79}, 026303 (2009).
\bibitem{FS01} M. Farge and K. Schneider, Coherent vortex simulation (CVS), a semi-deterministic turbulence model using wavelets, Flow, Turbul. Combust. {\bf 66}, 393 (2001).
\bibitem{SFPR05} K. Schneider, M. Farge, G. Pellegrino, and M. Rogers, Coherent vortex simulation of three-dimensional turbulent mixing layers using orthogonal wavelets, J. Fluid Mech. {\bf 534}, 39 (2005).
\bibitem{SV10} K. Schneider and O. V. Vasilyev, Wavelet methods in computational fluid dynamics, Annu. Rev. Fluid Mech. {\bf 42}, 473 (2010).
\bibitem{activematter} J. Urzay, A. Doostmohammadi, and J. M. Yeomans, Multi-scale statistics of turbulence motorized by active matter, J. Fluid Mech. {\bf 822}, 762 (2017). 
\bibitem{combust} J. Kim, M. Bassenne, C. A. Z. Towery, P. E. Hamlington, A. Y. Poludnenko, and J. Urzay, Spatially localized multi-scale energy transfer in turbulent premixed combustion, J. Fluid Mech. {\bf 848}, 78 (2018).
\bibitem{dropwave} A. Freund and A. Ferrante, Wavelet-spectral analysis of droplet-laden isotropic turbulence, J. Fluid Mech. {\bf 875}, 914 (2019). 
\bibitem{Morinishi1998} Y. Morinishi, T. S. Lund, O. V. Vasilyev, and P. Moin, Fully conservative higher order finite difference schemes for incompressible flow, J. Comput. Phys. {\bf 143}, 90 (1998).
\bibitem{Hirt&Cook1972} C. W. Hirt and J. L. Cook, Calculating three-dimensional flow around structures and over rough terrain, J. Comput. Phys. {\bf 10}, 324 (1972).
\bibitem{Onishi2011} R. Onishi, Y. Baba and K. Takahashi, Large-scale forcing with less communication in finite-difference simulations of steady isotropic turbulence, J. Comput. Phys. {\bf 230}, 4088 (2011).
\bibitem{Romain2010} R. Nguyen van yen, D. del-Castillo-Negrete, K. Schneider, M. Farge, and G. Chen, Wavelet-based density estimation for noise reduction in plasma simulations using particles, J. Comput. Phys. {\bf 229}, 2821 (2010).
\bibitem{Daubechies1992} I. Daubechies, {\it Ten lectures on wavelets} (Society for Industrial and Applied Mathematics, Philadelphia, 1992).
\bibitem{Mallat98} S. Mallat, {\it A wavelet tour of signal processing The Sparse Way, Third Edition} (Academic Press, Burlington, 2009).
\bibitem{SaitoGotoh2018} I. Saito and T. Gotoh, Turbulence and cloud droplets in cumulus clouds, New J. Phys. {\bf 20}, 023001 (2018).
\bibitem{Karp2019} K. Karpi\'nska, J. F. Bodenschatz, S. P. Malinowski, J. L. Nowak, S. Risius, T. Schmeissner, R. A. Shaw, H. Siebert, H. Xi, H. Xu, and E. Bodenschatz, Turbulence-induced cloud voids: observation and interpretation, Atmos. Chem. Phys. {\bf 19}, 4991 (2019).
\bibitem{TK2} T. Oujia, K. Matsuda, and K. Schneider, Divergence and convergence of inertial particles in high Reynolds number turbulence, 
arXiv:2005.00525 (2020).

\end{thebibliography}
\end{document}